\pdfoutput=1
\documentclass{aa}
\usepackage{txfonts}

\usepackage{amssymb,epsfig,color}
\usepackage{amsmath}
\usepackage{graphicx}
\graphicspath{{figures/}}

\usepackage{makecell}
\usepackage{siunitx}
\usepackage{booktabs}
\usepackage{multirow}
\usepackage{longtable}
\usepackage{lscape}
\usepackage[hyperindex,breaklinks=true, colorlinks,citecolor=blue,draft=False]{hyperref}
\usepackage{textcomp}
\newcommand{\micron}{\textmu m}

\usepackage{xcolor}
\usepackage{tikz}
\usetikzlibrary{positioning,arrows.meta}

\begin{document}

    \title{New cold brown dwarfs with SPHEREx}
    \subtitle{A blind search of the high-latitude sky for cold brown dwarfs:\\ 452 new discoveries and 296 first spectroscopic confirmations}
    \titlerunning{New cold brown dwarfs with SPHEREx}
    \authorrunning{R. Cepeda-Arroita \& R. Rebolo}

    \author{
      Roke Cepeda-Arroita\inst{1,2}\thanks{\email{roke.cepeda@iac.es}}
      \and
      Rafael Rebolo\inst{1,2}
    }

	\institute{
		Instituto de Astrof\'{i}sica de Canarias, 38200 La Laguna, Tenerife, Canary Islands, Spain
		\and
		Departamento de Astrof\'{i}sica, Universidad de La Laguna (ULL), 38206 La Laguna, Tenerife, Spain
	}

	\date{Received 27 November 2026 / Accepted ...}

\abstract{
A complete census of the coldest brown dwarfs, the late-T and Y dwarfs, is essential to understand the substellar population of the solar neighbourhood, but they have been found almost entirely through their WISE colours and proper motions, so the census misses those that move slowly or lie far away. SPHEREx now provides a low-resolution 0.75--5\,\micron\ spectrum of the whole sky, where these objects can be recognised whether or not they move. We present a blind spectroscopic search of SPHEREx for T5--Y dwarfs among all 708\,947 red WISE sources ($W1-W2\geq1.5$, $W2<16.5$) at $|b|>10^\circ$. The search rediscovers 92\% of the known T5--Y dwarfs in its selection (98\% at $W2<15$), with consistent types and proper motions. It yields 452 new T and Y dwarfs and the first spectroscopic confirmation of 296 photometric candidates. Of the new dwarfs, 15 are candidate Y dwarfs; since our fit also types some known late-T dwarfs as Y, we expect about six of them to be Y dwarfs and the others late-T dwarfs. The 425 new and 283 confirmed T5--Y dwarfs increase the number of known T5--Y dwarfs at $|b|>10^\circ$ to $W2=16.5$ by 104\%. The new dwarfs have model effective temperatures of 400--1350\,K and distances of 14--57\,pc (median 39\,pc); 34 lie within 25\,pc. Almost half of them show no significant motion: they are the slow members of the field population that motion surveys under-sample. Among them are slow-moving, nearby candidate Y dwarfs, 20 dwarfs faster than 80\,km\,s$^{-1}$, candidates for old members of the thick disk, and candidate wide binaries with Gaia stars. Together with the known ones they make up 62\% of the T5--Y1 dwarfs that a search with our completeness should find; the missing ones are faint ($W2=15$--16.5). All spectra and an atlas of every object are released with this paper.
}

	\keywords{
		brown dwarfs --
		stars: low-mass --
		solar neighborhood --
		infrared: stars --
		surveys --
		techniques: spectroscopic
	}

\maketitle

\section{Introduction}\label{sec:intro}

Brown dwarfs are too light to sustain hydrogen fusion; the more massive ones burn their deuterium early on, and all of them then cool throughout their lives. Since the first discoveries of brown dwarfs in the Pleiades and as companions to nearby stars \citep{Rebolo1995,Nakajima1995}, wide-area optical and infrared surveys have dramatically expanded the known population, revealing hundreds of objects spanning a broad range of masses, ages and temperatures; in particular, increasingly sensitive near- and mid-infrared surveys have enabled the identification of progressively cooler brown dwarfs. The coldest, classified as late-T and Y dwarfs ($T_\mathrm{eff}\lesssim1000$\,K, down to $\approx$285\,K for WISE~0855$-$0714; \citealt{Luhman2014,Luhman2024}), are the oldest or least massive. The number of brown dwarfs per spectral type in the solar neighbourhood records both the substellar mass function and the history of star formation in the Galactic disk \citep{Kirkpatrick2021,Kirkpatrick2024,Raghu2024}. An accurate determination requires a complete census, which remains incomplete for the coldest brown dwarfs even within 20\,pc \citep{Kirkpatrick2024,Meisner2026}.

Many of the known late-T and Y dwarfs were found with WISE \citep{Wright2010,Mainzer2014}, first by their mid-infrared colours \citep{Kirkpatrick2011,Cushing2011,Griffith2012} and then mainly by their proper motions \citep{Marocco2021,Schneider2026,Karpov2025,Brooks2024}. Colour alone cannot separate them from galaxies and reddened stars, which share the colours of T dwarfs \citep{Assef2018}; deep near-infrared imaging, now from Euclid \citep{Mellier2025}, helps for the fainter objects but over part of the sky. A systematic spectroscopic survey offers a means of overcoming these limitations by identifying brown dwarfs through their characteristic molecular absorption features, independently of their proper motions.

\begin{figure}[!htbp]
\centering
\includegraphics[width=\columnwidth]{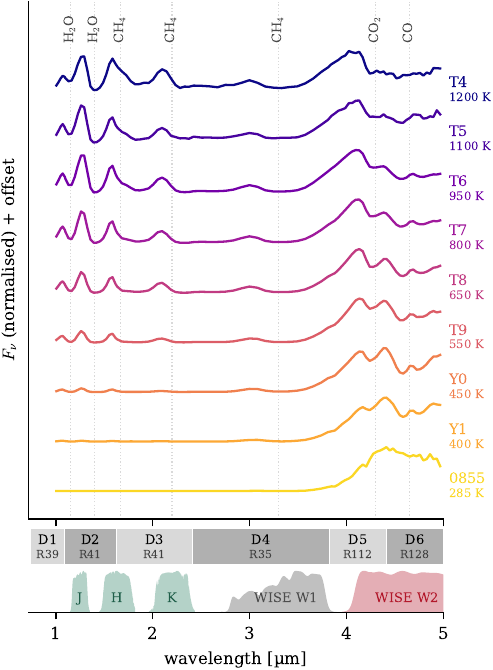}
\caption{What SPHEREx sees in a cold brown dwarf: the template spectra from T4 to Y1 (SPIFF hybrid templates: for T4 and T5 the mean SPHEREx spectrum of known dwarfs joined below 2.4\,\micron\ to a near-infrared spectral standard, for T6--Y1 JWST NIRSpec spectra) and of WISE~0855$-$0714, with the main molecular bands; next to each type, its typical effective temperature (median Elf Owl fit of the known dwarfs of that type, rounded to 50\,K; JWST value for WISE~0855). Below: the six SPHEREx detectors with their spectral resolving power $R$, the MKO $J$, $H$ and $K$ bandpasses (UKIDSS; VHS is similar) used in Sect.~\ref{sec:validation}, and the WISE W1 and W2 bandpasses. Cold brown dwarfs emit most of their light at 4--4.5\,\micron, which makes them bright in W2 and faint in W1.}
\label{fig:templates}
\end{figure}

Since 2025 SPHEREx \citep{Bock2025} has mapped the whole sky every six months with low-resolution spectra from 0.75 to 5\,\micron, a range that contains the most significant spectral features of a cold brown dwarf (Fig.~\ref{fig:templates}): the $J$ and $H$ peaks between water and methane bands, the deep methane band at 3.3\,\micron\ and the strong peak at 4--4.5\,\micron. Galaxies and reddened stars are unlikely to reproduce this combination of spectral features, making SPHEREx a promising tool for distinguishing cold brown dwarfs from potential contaminants, whether or not they move, provided they are bright enough. T dwarfs, with their $J$, $H$ and $K$ peaks, are the easiest to recognise. Y dwarfs emit most of their detectable flux at 4--5\,\micron, while their very faint emission at shorter wavelengths makes their molecular absorption features hard to detect, and hence harder to tell from other sources with red continua.

\citet{Gagne2026} demonstrated the potential of SPHEREx for the spectroscopic identification and characterisation of brown dwarfs. Extracting spectra for about 6000 ultracool dwarfs, most of them known objects or published candidates and 947 newly identified, they more than doubled the number of ultracool dwarfs with spectra known at the time and built the SPIFF library of SPHEREx templates that we use here; SPHEREx spectra of ultracool dwarfs have also been studied by \citet{Rustamkulov2026}, \citet{Tu2026}, \citet{BrooksCushingKothari2026} and \citet{Kiman2026}. These works start from objects that were already known or had been selected as candidates by their colours or motion. SPHEREx allows a different approach: a blind search of every red source in the sky, in which the spectrum alone can decide, and whose completeness can be measured. Such a search means measuring more than a hundred times as many spectra as SPIFF, the largest SPHEREx study of ultracool dwarfs so far.

In this paper we search the whole high-latitude sky for cold brown dwarfs with SPHEREx, without any motion requirement. We first describe the method (Sect.~\ref{sec:method}) and show that it works: it rediscovers most of the known T5--Y dwarfs (Sect.~\ref{sec:validation}). We then present the new objects (Sect.~\ref{sec:results}) and ask what they tell us about the census (Sect.~\ref{sec:science}).

\section{The method}\label{sec:method}

\begin{figure}[!htbp]\centering
\def\fsix{\fontsize{6.2}{7.2}\selectfont}%
\definecolor{cB}{HTML}{0072B2}\definecolor{cO}{HTML}{D55E00}\definecolor{cP}{HTML}{6A3D9A}\definecolor{cK}{HTML}{4D4D4D}
\resizebox{\columnwidth}{!}{%
\begin{tikzpicture}[font=\footnotesize, >={Stealth[length=1.4mm]},
  stage/.style={font=\fsix, text=#1, anchor=west},
  box/.style={draw=#1!65, fill=#1!5, rounded corners=2pt, align=center, inner sep=2.5pt, text width=70mm, execute at begin node={\hyphenpenalty=10000}},
  small/.style={draw=#1!65, fill=#1!5, rounded corners=2pt, align=center, inner sep=2.5pt, text width=20.6mm},
  arr/.style={->, draw=black!50, line width=0.55pt}]
\node[box=cB] (sel) {CatWISE2020 red sources at $|b|>10^\circ$:\\$W2<16.5$, $W1-W2\geq1.5$ or no $W1$\\[1pt]{\color{cB}\small 708\,947 targets}};
\node[box=cB, below=3mm of sel] (mea) {$\sim$350 SPHEREx exposures per target ($\sim$240 million cutouts); PSF fit with all neighbours; matched filter; SPHEREx position and motion; re-measure there\\[1pt]{\color{cB}\small 303\,711 detected, 116\,859 with a T4 or later template}};
\node[box=cO, below=3mm of mea] (vet) {physical score of the spectrum and position, threshold per type (T4--T8: fit vs blackbody and flat+power law, T flux ratios, flux at 0.75--1\,\micron, SPHEREx position offset; T9--Y: fit vs power law, flat+power law and M/L, methane, $J$, 4.5--5\,\micron, S/N, motion)\\[1pt]{\color{cO}\small 1690 candidates}};
\node[box=cO, below=3mm of vet] (eye) {inspected by eye: 1595 candidates (all but 36 saturated and 59 too faint), 2693 near misses below the thresholds and 137 objects from our target lists of known dwarfs and candidates; the eye sets the final labels, and the verdicts train the score};
\node[small=cK, below=3mm of eye.south west, anchor=north west] (det) {{\small 1412}\\Detections};
\node[small=cK, below=3mm of eye.south] (ten) {{\small 261}\\Tentative};
\node[small=cK, below=3mm of eye.south east, anchor=north east] (rej) {{\small 2752}\\rejected};
\node[box=cP, below=3mm of ten] (lit) {novelty check of every \mbox{inspected} object; of the Detections:\\[1pt]{\color{cO}\small 452 new}\quad{\color{cP}\small 296 first spectra}\quad{\color{cK}\small 652 known}};
\node[box=cK, below=3mm of lit] (sci) {types, temperatures, distances, velocities, census};
\draw[arr] (sel) -- (mea); \draw[arr] (mea) -- (vet); \draw[arr] (vet) -- (eye);
\foreach \b in {det,ten,rej} \draw[arr] (eye.south -| \b.north) -- (\b.north);
\draw[arr] (det.south) -- (det.south |- lit.north); \draw[arr] (lit) -- (sci);
\draw[arr] ([yshift=1.5mm]vet.east) -- ++(2.4mm,0) |- node[pos=0.25, right=-0.3mm, rotate=90, anchor=north, font=\fsix, text=black!60] {re-measure} (mea.east);
\draw[arr] (rej.east) -- ++(2.4mm,0) |- node[pos=0.25, right=-0.3mm, rotate=90, anchor=north, font=\fsix, text=black!60] {train the score} ([yshift=-1.5mm]vet.east);
\foreach \n/\t/\c in {sel/SELECT/cB, mea/MEASURE/cB, vet/VET/cO, eye/INSPECT/cO, det/CLASSIFY/cK, lit/NOVELTY/cP, sci/SCIENCE/cK} \node[stage=\c] at ([xshift=-15.5mm]sel.west |- \n) {\t};
\end{tikzpicture}}
\caption{The search, with the number of sources at each stage ($W2<16.5$, $|b|>10^\circ$; Sect.~\ref{sec:method}). We inspected the candidates and, to make sure that no good source was thrown away and to train the vetting, 2693 near misses below the thresholds, plus 137 objects from our target lists of known dwarfs and candidates. The Detection and Tentative counts include known dwarfs and objects later excluded (12 Detections earlier than T4, outside the selection or not brown dwarfs), so they exceed those of the catalogue (Table~\ref{tab:headline}). The known dwarfs, injected spectra and blank positions pass through the same chain and calibrate it (Sect.~\ref{sec:validation}).}
\label{fig:flow}
\end{figure}
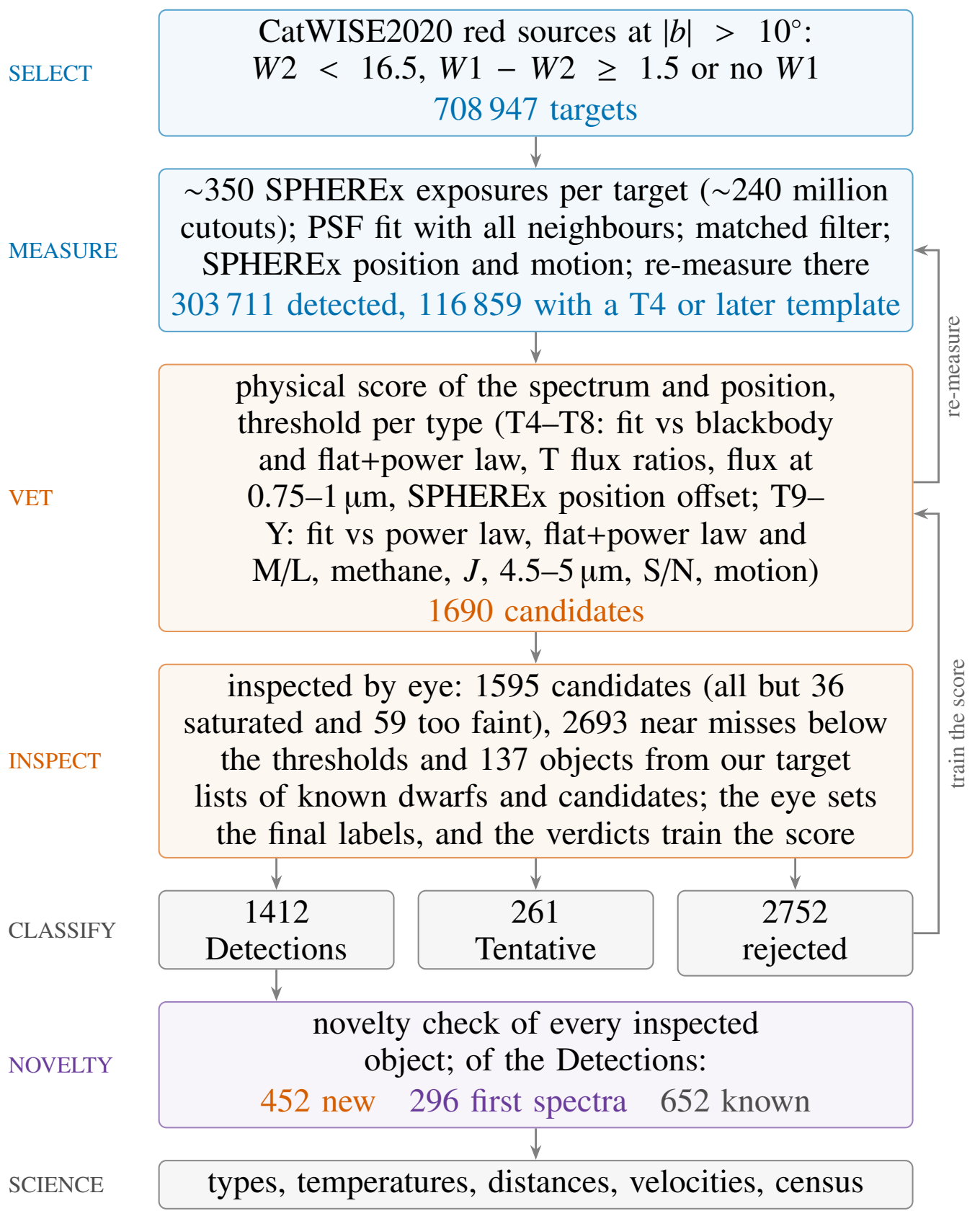

The method has three steps (Fig.~\ref{fig:flow}): we select every WISE source with a red $W1-W2$ colour, measure a SPHEREx spectrum for each, and keep those whose spectrum is that of a cold brown dwarf.

\subsection{Every red WISE source to $W2=16.5$}\label{sec:selection}

We search for dwarfs of type T5 and later, whose methane absorption at 3.3\,\micron\ makes them much redder in $W1-W2$ than almost any star. We therefore select from CatWISE2020 \citep{Marocco2021}, which is 90\% complete at $W2\approx17.5$, more than a magnitude deeper than SPHEREx reaches, every source with
\begin{equation}
W2<16.5,\quad |b|>10^\circ,\quad W1-W2\geq1.5\ \mbox{or no W1 detection},
\label{eq:selection}
\end{equation}
detected in W2 at S/N\,$\geq5$ with clean artefact flags. The colour limit keeps 85\% of the known T5--T6 dwarfs and 96--99\% of the later ones. There is no requirement on proper motion. This gives 708\,947 targets, most of them galaxies and reddened stars, all measured, vetted and inspected in the same way.

\subsection{A SPHEREx spectrum for every target}\label{sec:measurement}

SPHEREx maps the whole sky every six months. Its linear variable filters record a different wavelength in each exposure of a given position, and six detectors cover 0.75--5\,\micron\ in 102 spectral channels ($R\approx35$--130) with 6.15\arcsec\ pixels. Our data are the Level~2 images taken between 24 April 2025 and 5 October 2026 and served by IRSA (quick releases QR2 and QR3); they cover most targets in three all-sky passes, with a median of 350 exposures per target (16th--84th percentiles 230--410). The targets fall on about a million Level~2 images, some 70\,TB of data. From them we read about 240 million exposure cutouts of the targets, and many more for the blank positions described below, each fitted separately, in about 140\,000 CPU hours on the IAC computing cluster.

\begin{figure}[!htbp]
\centering
\includegraphics[width=\columnwidth]{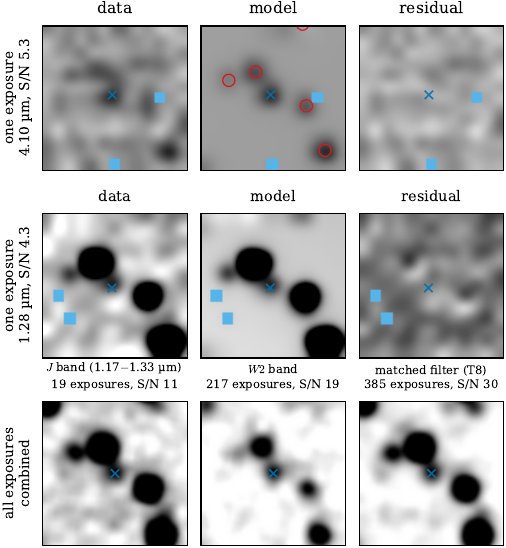}
\caption{The forced photometry of CWISE~J043050.49$-$072453.9, a new T8 dwarf with $W2=15.5$ (Fig.~\ref{fig:example}). Top two rows: one exposure at the 4.1\,\micron\ peak and one at the $J$ peak (1.28\,\micron), in the $13\times13$-pixel fit box ($80''$ across) and in units of the pixel noise: the data, the model (a planar background, the PSFs of all CatWISE2020 and Gaia neighbours, red circles, and of the target, blue cross) and the residual; the light-blue squares are pixels flagged in the Level~2 mask, left out of the fit. A single exposure detects the target at only S/N\,$\approx4$--5. Bottom row: all the exposures combined on the sky (north up, east left; neighbours not subtracted) in the $J$ band, in the $W2$ band and with the T8 template over 0.75--5\,\micron\ (the template fit of Sect.~\ref{sec:measurement}), with the number of exposures and the formal S/N of their combined PSF fits above each panel: together, the 19 $J$ exposures detect the target at S/N\,=\,11. The template fit counts the exposures with valid fluxes where the template is not zero, and its S/N exceeds that of the 3.8--5\,\micron\ fit of Fig.~\ref{fig:example}f (23) because it also uses the $J$ and $H$ peaks.}
\label{fig:fitdemo}
\end{figure}

\begin{figure*}[!tbp]
\centering
\includegraphics[width=0.95\textwidth]{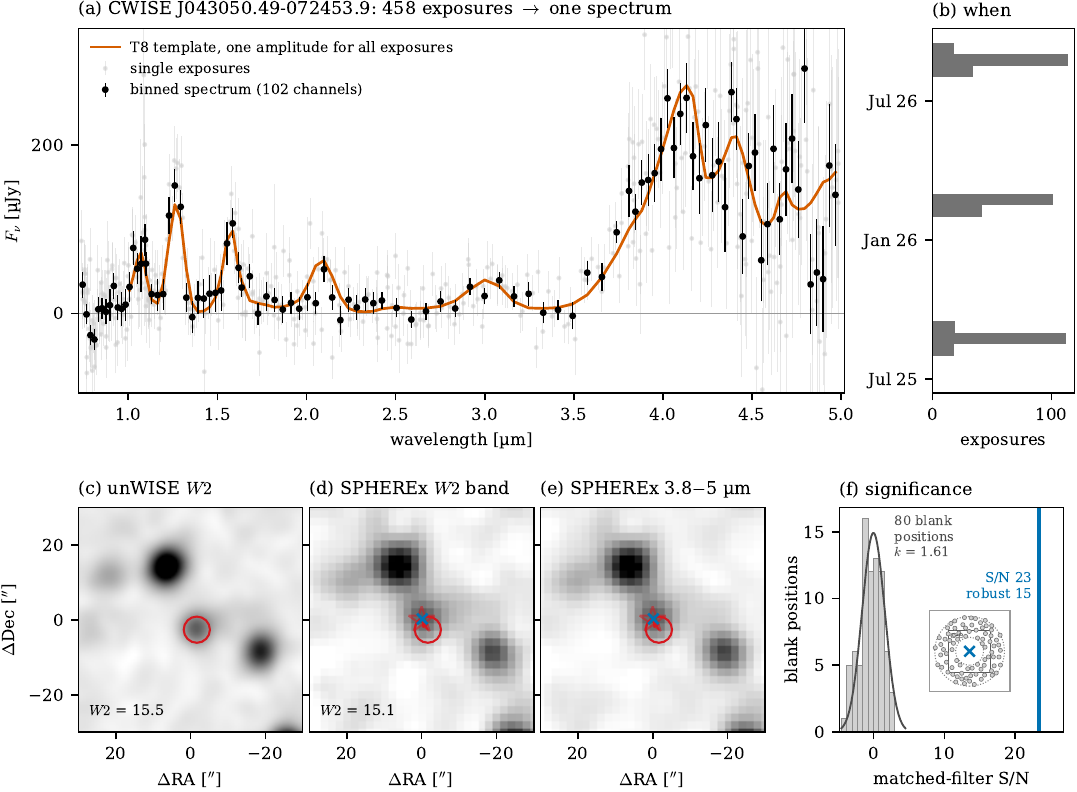}
\caption{How the search sees one of the new T dwarfs, CWISE~J043050.49$-$072453.9 (T8). (a) Its SPHEREx spectrum: the fluxes measured in the 458 individual exposures (grey, with their errors), binned into 102 channels (black), and the best template (orange), with one amplitude fitted to all exposures. (b) When the exposures were taken. (c) The unWISE $W2$ image \citep{Lang2014} with the CatWISE2020 magnitude, (d) the SPHEREx image in the WISE $W2$ band (each exposure weighted by the $W2$ response), with the $W2$ magnitude synthesised from the spectrum, and (e) the SPHEREx image at 3.8--5\,\micron\ (detectors D5--D6, each exposure weighted by the T8 template: the matched filter used for detection); the three share the same axes, in the order of the atlas pages. The red circle marks the CatWISE2020 position, the hollow red star where its CatWISE2020 proper motion puts it at the SPHEREx epoch (4.5\arcsec\ away), and the blue cross where SPHEREx finds the source: on the predicted position. (f) The significance: the matched-filter S/N at 80 blank positions 20--50\arcsec\ from the target (inset; dotted circles at 20 and 50\arcsec, square: the field of panels c--e) scatters by $k=1.61$ times the formal value (curve), which turns the formal S/N of 23 into a robust S/N of 15. The online atlas (Sect.~\ref{sec:conclusions}) gives one page for every known, new and tentative object, with its spectrum, the unWISE and SPHEREx matched-filter images (also with the catalogued neighbours subtracted), the proper motion and its significance, and the measurements of Appendix~\ref{app:tables} (Appendix~\ref{app:atlas}).}
\label{fig:example}
\end{figure*}

For every exposure we read a $31\times31$-pixel cutout from the Level~2 image \citep{Akeson2025} and fit its central $13\times13$ pixels, by weighted least squares, with the PSF of that exposure at the target's predicted position (its CatWISE2020 position moved with the CatWISE2020 proper motion to the date of the exposure; we call it the predicted position throughout), the PSFs of all CatWISE2020 and Gaia DR3 \citep{GaiaMission2016,GaiaDR3} neighbours in the cutout, and a planar background. Each exposure thus gives the flux of the target with its neighbours removed (Fig.~\ref{fig:fitdemo}; Fig.~\ref{fig:example}a, grey points). The fluxes are kept as measured, negative values included (there is no positivity prior), so that the noise of a faint source averages out instead of biasing its spectrum.

A cold dwarf at $W2\approx15.5$ has a S/N of only 1--3 in a single spectral channel, so we never fit the channels separately. Instead we fit a single template amplitude to all exposures at once, with the template evaluated at the wavelength of each exposure and the neighbours and background fitted exposure by exposure: a matched filter, the optimal combination of all the exposures for detecting a source of known spectral shape. Because it assumes that shape, the search is most sensitive to dwarfs whose spectra resemble our templates, and less so to peculiar ones, such as metal-poor subdwarfs. We use two such fits. The best template of the bank (Fig.~\ref{fig:templates}) over 0.75--5\,\micron\ gives the spectral type and, for T dwarfs, gains from the $J$ and $H$ peaks; a fit restricted to 3.8--5\,\micron, where cold dwarfs are brightest and most other sources faint, is the most sensitive for Y dwarfs. The binned spectrum (Fig.~\ref{fig:example}a), the weighted mean of the exposures in each of the 102 channels, serves for the spectral-shape tests and to refine the types. We show spectra as $F_\nu$ in $\mu$Jy, the natural unit of the SPHEREx photometry; the released spectra are also given in $F_\lambda$.

The flux errors of each exposure come from the variance image of the Level~2 data, propagated through the least-squares PSF fit of that exposure; we scale them for each detector by the square root of the median reduced $\chi^2$ of the fits, so that they describe the actual scatter. The matched filter combines the exposures with inverse-variance weights, and its formal S/N is the fitted amplitude divided by its propagated error. Confusion with faint uncatalogued sources makes this formal S/N optimistic. We therefore repeat the 3.8--5\,\micron\ fit at 80 blank positions around the target and divide the formal S/N by the scatter $k$ of these values to obtain the \emph{robust} S/N (Fig.~\ref{fig:example}f). The positions are drawn at random, uniformly in area, in the annulus 20--50\arcsec\ from the target, keeping only positions at least one SPHEREx pixel (6.15\arcsec) from all those already chosen; the annulus holds about 174 such independent positions. They thus sample the same field, with the same crowding, depth and background as the target, without any further selection. Each blank position is measured exactly like the target: the same exposures (at least 10), the same template and error scaling, the position moving with the target's catalogued motion, and all catalogued neighbours (CatWISE2020 and Gaia) and the target itself modelled as neighbours. $k$ is the standard deviation of the 80 values after rejecting those more than four median absolute deviations from their median, so that a real uncatalogued source at one position does not inflate it; with 80 positions it is reproducible to about 6\% (two independent sets of positions for 400 targets). 

We also measure the position of each target in the SPHEREx data. We repeat the 3.8--5\,\micron\ fit, with all catalogued neighbours fitted at the same time, with the target at trial positions 1\arcsec\ apart, up to 6\arcsec\ (one SPHEREx pixel) from the predicted position in each coordinate. The trial position with the highest S/N is then refined continuously to the minimum of the $\chi^2$, and the curvature of the $\chi^2$ gives its uncertainty. The grid only provides a starting point: because the fit combines hundreds of exposures, each with its own PSF, the $\chi^2$ minimum locates the source to a fraction of the 6.15\arcsec\ pixel, 0.4--0.7\arcsec\ for the S/N of 10--20 typical of our candidates, and bootstrapping the exposures gives the uncertainty of each position (calibrated on the known dwarfs, Sect.~\ref{sec:val_pm}). Compared with the CatWISE2020 position, which refers to the mean epoch of the WISE observations of each source (2014.8--2015.9 for 95\% of the Detections), it gives the proper motion over the $\approx$10.5\,yr to the mean SPHEREx epoch of each target, to $\approx$40--70\,mas\,yr$^{-1}$, comparable to the CatWISE2020 motions themselves. Where the SPHEREx position lies more than 1\arcsec\ from the prediction, usually because the catalogue motion is inaccurate, we measure the spectrum again there, unless a catalogued neighbour of similar brightness lies within 3\arcsec\ of both positions. Only a dwarf whose CatWISE2020 motion is wrong by more than 0.6\arcsec\,yr$^{-1}$ would fall outside the searched area. Of the known dwarfs in the selection that move faster than 1\arcsec\,yr$^{-1}$, we recover 38 of the 40 with literature motions (95\%, against 91\% below 0.3\arcsec\,yr$^{-1}$; Sect.~\ref{sec:val_known}): 31 of the 32 at $W2<15$, including the fastest, WISE~J163940.83$-$684738.6 and 2MASS~J11145133$-$2618235 (3.2 and 3.0\arcsec\,yr$^{-1}$), and 7 of the 8 at $W2=15$--16.5 (against 76 of the 92 slower than 0.3\arcsec\,yr$^{-1}$ in that range); the two others fall below the detection threshold. The search does need a CatWISE2020 entry with a usable motion, which the fastest movers may lack: the CatWISE2020 coadds span eight years, over which WISE~0855$-$0714 (8.1\arcsec\,yr$^{-1}$; Y4, beyond our T5--Y1 sample) moves by more than 1\arcmin, and the catalogue lists it only as two faint fragments ($W2=16.4$--16.5) near its 2015 position, where SPHEREx finds nothing. Measured at its true position, it is detected at a matched-filter S/N of 35 (16 after the blank-position scaling), with a SPHEREx $W2=13.88\pm0.03$. A dwarf moving faster than a few arcseconds per year, and so within a few parsecs, could therefore be missed if CatWISE2020 fragments it; a search around the positions of such fragments is left for future work.

\subsection{Vetting}\label{sec:vetting}

Most targets are galaxies, reddened stars or too faint to be detected. The vetting reduces them to the few thousand that look enough like a cold dwarf to be worth inspecting by eye, losing as few real dwarfs as possible and none systematically; it does not set the final labels. It has three steps: detection, ranking and exclusion.

\paragraph{Detection.} A target is detected if its robust S/N at 3.8--5\,\micron\ is at least 4.5, or its template S/N over 0.75--5\,\micron, divided by the same factor $k$ (median $k=1.74$), is at least 6. Because $k$ is floored at 1, the robust S/N never exceeds the formal one, so a target whose formal S/N misses both thresholds is undetected whatever its $k$; we measure $k$ with the 80 blank positions for every target whose formal S/N reaches them (72\% of the targets). Most targets that are not detected are simply too faint: the undetected fraction rises from 5\% at $W2<15$ to 14\% at $W2=15$--15.5, 41\% at 15.5--16 and 83\% at 16--16.5.

\paragraph{Ranking.} About 117\,000 targets are detected with a T4 or later best template, almost all of them galaxies, reddened stars or blends. The vetting ranks them with a \emph{physical score}, a sum of one monotonic function per measured quantity fitted to our visual verdicts, with a threshold per type (Appendix~\ref{app:score}). For T4--T8 templates five quantities suffice: how much better the template fits than a blackbody and than a flat spectrum to 3\,\micron\ joined to a power law beyond (the shape of the most common contaminants), the T-dwarf flux ratios, the flux at 0.75--1\,\micron\ and the offset of the SPHEREx position. T9--Y spectra differ less from a red continuum, and their score uses eight quantities, including the methane trough, the $J$ peak and the motion. Vetting and inspection iterated: each round of verdicts refitted the score, and targets just below the thresholds were inspected too. On objects not used in the fit, the thresholds keep 94--98\% of the Detections of each type and admit at most 0.2\% of random targets.

\paragraph{Exclusion arguments.} Three tests are decisive where they apply, but apply to few targets, so they are not part of the score; a candidate that fails one is rejected unless the inspection shows otherwise: a 12\,\micron\ detection with $W2-W3>3.5$ at S/N\,$\geq5$ (dusty galaxies, AGN and young stars; only one of 455 known T5--Y dwarfs with an AllWISE counterpart has even $W2-W3>3$ at $W3$ S/N\,$\geq3$), a Gaia DR3 source within 1.5\arcsec\ at $W2>14$ (a T dwarf this faint is far below the Gaia limit), and a $z$-band detection with $z-W2<4.5$, bluer than any known T dwarf. A SPHEREx position on another catalogued source is also rejected. A $z$ counterpart with $z-W2=4.5$--6.5 rules out a T9 or Y dwarf but not a T5--T8 one, so it only flags the type.

\paragraph{Inspection.} No automatic measurement catches every failure. We met a bright neighbour pulling the fit; an artefact, such as an unflagged cosmic ray, in a few exposures; WISE entries that misplace the source and so mimic a motion (a position pulled towards a nearby star, a catalogued motion of zero for a source that moves); optical ghosts of bright stars; and fast movers with several CatWISE2020 entries along their path, which, fitted as neighbours, absorb part of the target's flux. A vetting that rewards motion selects such cases preferentially. We therefore inspected every candidate, except the saturated ones and a few dozen too faint to judge (Appendix~\ref{app:score}), on an atlas page like Fig.~\ref{fig:example} and classified it as a \emph{Detection}, a \emph{Tentative} object or rejected; the inspection sets the final labels. The detection thresholds use one number, the robust S/N; the inspection can use more. An object just below the threshold whose $J$, $H$ and $K$ peaks fall exactly at the template wavelengths, or that moves significantly, is as secure as one above it, and we keep it as a Detection. For Y spectra, nearly featureless at our S/N, a Detection also requires a motion confirmed by eye; T9 spectra keep enough methane and $J$-band structure to be judged on the spectrum. Tentative objects have plausible but weaker spectra, or Y spectra without that motion; they are listed but not used in any statistic.

\section{A blind rediscovery of the known T dwarfs}\label{sec:validation}

\subsection{The known T5--Y brown dwarfs with spectra}\label{sec:val_known}

\begin{figure*}[!tbp]
\centering
\includegraphics[width=\textwidth]{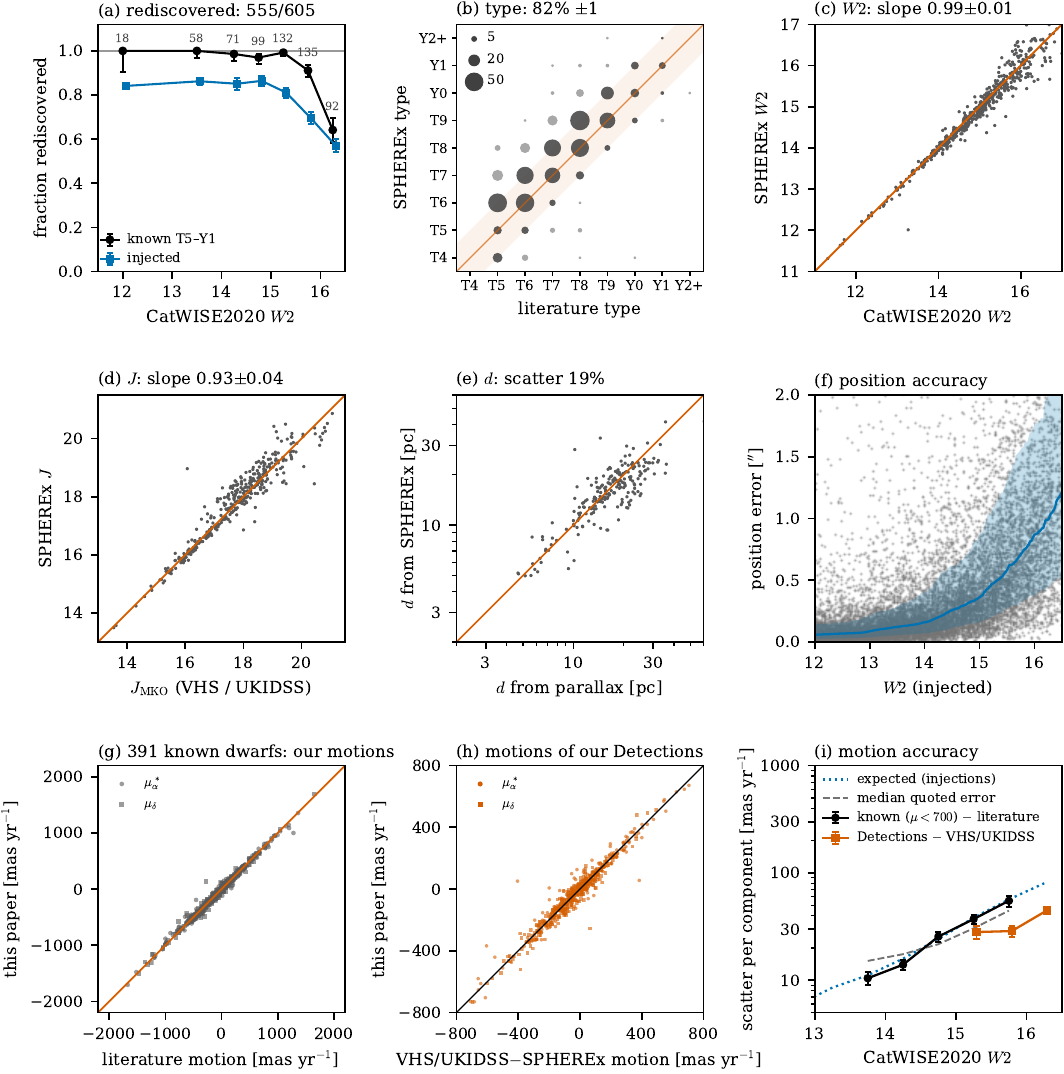}
\caption{Validation of the method: detection, types, photometry, distances, positions and motions. (a) Fraction of the known T5--Y1 dwarfs in the selection recovered by the blind search, i.e.\ detected and above the vetting thresholds (black; numbers: known dwarfs in the selection per bin, 605 in all, of which 555 are recovered), and of the injected spectra (blue; weighted by type like the known sample), against $W2$, with 1$\sigma$ errors (Sect.~\ref{sec:val_inj}). (b) SPHEREx against literature spectral type: number of objects in each cell (circle area; dark grey within $\pm$1 subtype, light grey beyond); orange line: equality, shaded band: $\pm$1 subtype. (c) SPHEREx synthetic $W2$ against CatWISE2020 $W2$ and (d) SPHEREx synthetic $J$ against $J_{\rm MKO}$ from VHS and UKIDSS; orange: equality; the slope in the title is from a weighted linear fit. (e) Spectrophotometric distance from the SPHEREx type and $W2$, as used for the new objects, against the parallax distance, for the known T5--Y dwarfs with parallaxes better than 10\%; the title gives the scatter. (f) Positional accuracy from the injections (moving with the motions of real dwarfs, $W2$ drawn uniformly): distance between the recovered and the true injected position at the SPHEREx epoch against $W2$, for the injected T and Y spectra detected at S/N\,$\geq5$ (grey points); blue: running median and 16--84\% range (13\,179 injections). (g) Our motions (CatWISE2020 to SPHEREx) of the 391 known T5--Y dwarfs against their literature proper motions (circles: $\mu_\alpha^*$, squares: $\mu_\delta$); orange: equality. (h) Our motions of the new and confirmed Detections against the independent motions from VHS/UKIDSS to SPHEREx (orange; circles and squares as in g; black: equality). (i) Accuracy of the motions per component against $W2$, in bins of 0.5\,mag (errors from bootstrap resampling of the objects): scatter of our motions relative to the literature motions for the 319 known dwarfs of (g) slower than 700\,mas\,yr$^{-1}$ (black) and relative to the VHS/UKIDSS motions of the Detections in (h) (orange; smaller, because the $J$-band positions are more precise than the WISE ones and the baseline is longer, 13\,yr); median uncertainty per component quoted for our motions (dashed); and the expectation from the position errors of the injections in (f), divided by the $\approx$10.5-yr baseline (dotted).}
\label{fig:validation}
\end{figure*}

We count a dwarf as known only if it has a published spectroscopic type, in the UltracoolSheet \citep{UltracoolSheet}, the SPIFF compilation of SPHEREx spectra \citep{Gagne2026} or \citet{Burgasser2025}. Dwarfs whose published types are photometric only are counted as candidates, although the UltracoolSheet lists some of them as known, and our spectra are their first: the Backyard Worlds discoveries of \citet{Schneider2026} without a SPIFF type, the types from Spitzer colours of \citet{Meisner2020a,Meisner2020}, the methane-imaging types of \citet{Cardoso2015} and \citet{Tinney2018}, and the types estimated by \citet{Kirkpatrick2021}. We test the method on the known T5--Y1 dwarfs: the 675 with $W2<16.5$ at $|b|>10^\circ$ in the UltracoolSheet or SPIFF, merged and de-duplicated, including 87 Backyard Worlds discoveries that SPIFF types. The same sample is used throughout the paper. The blind search finds them exactly as it finds any other object, from their CatWISE2020 entries and with the same measurement, thresholds, vetting and inspection. Of the 605 that meet our selection (most of the others are T5--T6 dwarfs with $W1-W2<1.5$), the measurement and vetting recover 555 (92\%), nearly all of them to $W2=15.5$ and 64\% at 16--16.5, where the S/N is low (Table~\ref{tab:val}; Fig.~\ref{fig:validation}a). The known dwarfs that the search misses are mostly faint, with low S/N and no significant motion; the fast movers are recovered as often as the others (38 of the 40 faster than 1\arcsec\,yr$^{-1}$; Sect.~\ref{sec:measurement}). We also processed the 70 known dwarfs outside the selection, to follow the completeness beyond it.

The SPHEREx types agree with the literature within one subtype for 82\% of the objects (Fig.~\ref{fig:validation}b). At a given literature type they are on average 0.3 subtypes later, mostly because dwarfs typed T5 in the literature are fitted best by our T6 template; conversely, at a given SPHEREx type the literature types are on average 0.5--1.2 subtypes earlier from T6 to Y1, so the SPHEREx types are not on the near-infrared scale and run late. Where they differ, the reason is mostly the wavelength range: most literature types rest on 1--2.5\,\micron\ spectra, at higher resolution but without the 3.5--5\,\micron\ region that SPHEREx adds, where late-T and Y dwarfs emit most of their flux and which is the most sensitive to temperature. The types are least certain at the cold end: at our S/N the T9, Y0 and Y1 templates differ mainly in the shape of the 4--5\,\micron\ peak, so the types of T9--Y dwarfs are uncertain by one to two subtypes. Of the known dwarfs that SPHEREx types Y0, 70\% are T9.5 or earlier in the literature, and of those typed Y1, 21\%, and for most of them with parallaxes the absolute $W2$ magnitude agrees better with the literature type. Applied to our new candidate Y dwarfs (12 Y0 and 3 Y1; Sect.~\ref{sec:props}), these fractions predict that about six are Y dwarfs and the others T8--T9 dwarfs.

A residual background can shift the types too: in crowded fields, the residuals of bright or uncatalogued neighbours add a nearly constant flux to a spectrum. Such an offset matters only where it is comparable to the flux of the dwarf, that is, for the faintest objects: at the median $W2$ of the new dwarfs (16.1), an offset of $-16\,\mu$Jy, the median of those we fit below, is a fifth of the 4\,\micron\ peak flux, whereas for the known dwarfs, about a magnitude brighter (median $W2=15.1$), it is less than a tenth. Fitting an offset for every object therefore barely changes the types of the known dwarfs (the fraction within one subtype of the literature goes from 82\% to 79\%), and it can absorb genuine differences in spectral shape. We fit it only where it corrects a measurable error: where it is significant and negative ($\Delta\chi^2>9$) and the flux in the absorption troughs (0.75--0.97, 1.38--1.50, 1.80--2.00 and 2.4--3.2\,\micron), where the templates predict almost none, lies below the template fitted without an offset (by more than $1.5\sigma$ on average). Such spectra sit below zero while their $J$, $H$ and $K$ peaks are clearly present, and without the offset they are typed too late, as Y0 rather than T9, for example. The criterion selects 68 catalogue objects (54 Detections and 14 Tentative ones), most of which move one or two subtypes earlier, and 42 known dwarfs, of which 36 then agree with their literature types within one subtype, against 29 without the offset. It also brings the SPHEREx $W2$ of these objects into agreement with CatWISE2020 (median difference $+0.16$\,mag without the offset and $-0.10$\,mag with it). The atlas shows their spectra with the offset removed.

We therefore use the SPHEREx types as measured, and give for every object the range of templates within $\Delta\chi^2\leq1$ of the best. These types illustrate the result of an automatic fit and indicate the likely spectral class of each object; they do not replace near-infrared spectral typing, which requires follow-up spectroscopy, especially at low S/N. Because the latest and faintest objects are uncertain by up to two subtypes, the tables list every object of SPHEREx type Y as T9--Y (except the peculiar CWISE~J001146.08$-$471306.6, Sect.~\ref{sec:props}). For the distances we calibrate the absolute magnitude of each SPHEREx type on the known dwarfs with parallaxes better than 10\% (for T5, which few known dwarfs with parallaxes have as their SPHEREx type, we keep the literature value; Fig.~\ref{fig:validation}e): because the absolute magnitudes are calibrated on the SPHEREx types themselves, the difference between the SPHEREx and the literature types does not bias the distances.

The SPHEREx photometry reproduces the survey magnitudes over the whole range (Fig.~\ref{fig:validation}c,d): weighted linear fits give slopes of $0.99\pm0.01$ against CatWISE2020 $W2$ and $0.93\pm0.04$ against the $J$ magnitudes of VHS and UKIDSS, and the zero points agree to a few hundredths of a magnitude. SPHEREx $W2$ is $0.05\pm0.01$\,mag brighter than CatWISE2020, mostly because our synthetic $W2$ averages the spectrum over the band only to 5.0\,\micron\ (85\% of the response) and so leaves out its red end, where cold dwarfs are fainter than the band average, which raises the average (0.03\,mag for the JWST spectrum of WISE~0855$-$0714). The near-infrared colours confirm the nature of the Detections: for the 415 T5--Y Detections with a VHS, UKIDSS or UHS $J$ magnitude (48 of them T9--Y), $J-W2$ lies within the range of the known dwarfs of the same SPHEREx type for 389, and within that of an adjacent type, or 0.12\,mag of it, for all but two: CWISE~J084307.85$-$121605.9 (T7, $J-W2=1.1$) and the Y pec dwarf CWISE~J001146.08$-$471306.6. The values, 1.1--4.9, are far from the extremes of the cold-dwarf sequence ($J-W2\approx11.6$ for WISE~0855$-$0714; \citealt{Leggett2025}). The $J$-detected T9--Y dwarfs lie on the blue side of the sequence, as expected near the $J$ detection limits ($J\lesssim20$--21).

\begin{table}[!htbp]
\caption{Blind rediscovery of the known T5--Y1 dwarfs ($W2<16.5$, $|b|>10^\circ$). The known sample includes the dwarfs bluer than our colour cut ($W1-W2<1.5$), which lie outside the selection.}
\label{tab:val}
\centering\footnotesize
\sbox0{\setlength{\tabcolsep}{6.4pt}\begin{tabular}{lcccc}
\toprule
$W2$ (mag) & Known$^{a}$ & In selection & Rediscovered$^{b}$ & Fraction \\
\midrule
$<15$ & 264 & 246 & 242 & 98\% \\
15--15.5 & 147 & 132 & 131 & 99\% \\
15.5--16 & 150 & 135 & 123 & 91\% \\
16--16.5 & 114 & 92 & 59 & 64\% \\
\midrule
$<16.5$ & 675 & 605 & 555 & 92\% \\
\bottomrule
\end{tabular}}\usebox0\par\vspace{2pt}
\parbox{\wd0}{\raggedright\scriptsize $^{a}$With a published spectroscopic type (UltracoolSheet and SPIFF, merged); dwarfs with only photometric types (Sect.~\ref{sec:val_known}) are not included; we also exclude ULAS~J1319+1209 (proper motion refuted) and ULAS~J1307+3325 (M-type spectrum), listed as late-T dwarfs but not brown dwarfs. $^{b}$Detected and above the vetting thresholds in the blind search.}
\end{table}

\subsection{Injected spectra and blank sky}\label{sec:val_inj}

The known objects are a convenient test, but not a random one: earlier surveys found them, and almost all of them move. To measure the completeness for the objects we are looking for, we injected 2347 real SPHEREx spectra of known T5--Y1 dwarfs, scaled to $W2=11$--16.5, into the stored SPHEREx exposures of randomly chosen search targets, 25--37\arcsec\ from the target: in every exposure we added the PSF of that exposure, scaled to the flux of the spectrum at its wavelength, so the simulated dwarfs carry the real noise and neighbours of the data. Each simulated dwarf moves with the proper motion of a known dwarf of similar type and brightness, and its catalogue entry carries the CatWISE2020 measurement of that motion, as for the search targets. We processed them blindly, with the same detection thresholds and vetting. Weighted by type like the known sample, the search recovers 85\% of them at $W2<15$, 81\% at 15--15.5, 69\% at 15.5--16 and 57\% at 16--16.5 (Fig.~\ref{fig:validation}a), 7--22 percentage points less than it rediscovers of the known dwarfs, which lie in cleaner fields than average. The coldest types are the hardest: at $W2=15.5$--16 the recovery is 78\% for T5--T6, 73\% for T7--T8 and 46\% for T9--Y1. We therefore use the injections, by type and magnitude, as the completeness of the search (Sect.~\ref{sec:census}). They pass through the measurement and the vetting but not the inspection, which rejects few real dwarfs: of the known dwarfs above the vetting thresholds that we inspected, nearly all are Detections. The purity of the inspected candidates is set by the inspection itself; of the inspected candidates of the final vetting, about 70\% are Detections.

Finally, we ran the whole method on 6000 empty positions, with no injected source, 60--120\arcsec\ from randomly chosen search targets; each position was given the catalogue magnitude and motion of its target, as if a source were there. Eleven pass the vetting thresholds, all with a T9 or Y template; eight of them land on a catalogued source and are rejected by that rule. Noise and confusion alone thus send about $5\times10^{-4}$ of the targets to inspection, all as T9--Y candidates. A real T9 dwarf is recognised by its methane and $J$-band structure, but a Y spectrum at low S/N is not, which is why a Y Detection also requires a motion confirmed by eye.

\subsection{Positions and proper motions}\label{sec:val_pm}

Our proper motions are weighted least-squares linear fits to the CatWISE2020 position, which refers to the mean epoch of the WISE observations of the source (not to the 2015.4 reference epoch of the CatWISE2020 motion fit), the SPHEREx position, fitted directly to the exposures (Sect.~\ref{sec:measurement}), and, for the 59\% of the Detections found there, the $J$-band position from VHS DR5 \citep{McMahon2013}, UKIDSS LAS \citep{Lawrence2007} or UHS \citep{Dye2018}, which extends the baseline from $\approx$10.5 to $\approx$13\,yr. Each position has its own uncertainty: the bootstrap error of the SPHEREx position (resampling the exposures) plus a term that grows towards low S/N, the catalogue error of the CatWISE2020 position, scaled, and for the $J$-band position a term that grows with $J$, all calibrated below.

We calibrate these uncertainties on the known dwarfs. The bootstrap alone underestimates the scatter about the literature motions (normalised robust scatter 1.5), because it cannot see errors common to all exposures (point-spread-function and astrometric-model errors, faint blended sources), and the catalogue uncertainties of the CatWISE2020 positions are themselves too small \citep{Marocco2021}. We therefore add in quadrature to the bootstrap error of each SPHEREx position a term $\sqrt{a^2+(b/{\rm S/N})^2}$, multiply the CatWISE2020 catalogue errors by a factor $k$, and give the $J$-band positions an error $\sqrt{c^2+(d\,10^{0.4(J-18)})^2}$; a maximum-likelihood fit to 448 known T5--Y dwarfs with literature motions, 233 of them with a $J$-band position, gives $a=0.13\pm0.02\arcsec$, $b=2.0^{+0.5}_{-1.3}$\arcsec, $k=1.58\pm0.14$, $c=0.12\pm0.01\arcsec$ and $d=0.03\pm0.01\arcsec$. The differences from the literature, divided by the combined uncertainties, then have a robust scatter of 0.97 (1.01 for the dwarfs with a $J$-band position), between 0.83 and 1.09 in every bin of $W2$ and S/N except the faintest one in S/N (0.77, where the errors are slightly overestimated), with 3\% beyond $3\sigma$, and the SPHEREx position errors agree within 20\% with those measured on the injected sources. Adding the $J$-band position reduces the motion uncertainty of those objects by a quarter (median 39 instead of 53\,mas\,yr$^{-1}$ for the new Detections); we drop it for the 13 objects whose three positions are inconsistent with a straight line.

Three independent tests show that motions and uncertainties are reliable (Fig.~\ref{fig:validation}f--i). First, injected spectra that move like real dwarfs are recovered within 0.1\arcsec\ (median) of their true position at $W2<14$ and within 1.0\arcsec\ at $W2=16$--16.5 (Fig.~\ref{fig:validation}f). Second, for 391 known T5--Y dwarfs our motions follow the literature, with slopes of $1.00\pm0.01$ below and $0.977\pm0.007$ above 0.7\arcsec\,yr$^{-1}$ (Fig.~\ref{fig:validation}g); this requires the CatWISE2020 position at its own epoch, which for the fastest dwarfs is on average 0.6\,yr later than 2015.4 (taking 2015.4 would make their motions 6--8\% too low). Third, a second motion, from the $J$-band position to SPHEREx over $\approx$13\,yr, reproduces the literature motions of the known dwarfs and agrees with our CatWISE2020-to-SPHEREx motions of the Detections without bias: their difference, divided by its uncertainty (which accounts for the shared SPHEREx position), has a robust scatter of 0.96 (Fig.~\ref{fig:validation}h,i). Panels g--i therefore show the two-position motions, independent of the $J$-band data used in the catalogue motions.
The accuracy per component thus grows from 10--14\,mas\,yr$^{-1}$ at $W2=13.5$--14.5 to $\approx$45\,mas\,yr$^{-1}$ for the Detections at $W2=16$--16.5, following the positional accuracy (Fig.~\ref{fig:validation}i). We list a motion only when it is significant at $\geq3\sigma$ (Tables~\ref{tab:fullnew} and \ref{tab:fullconf}), which at $W2\approx15.7$ means faster than $\approx$130\,mas\,yr$^{-1}$.

\section{New T and Y dwarfs}\label{sec:results}

\subsection{452 new discoveries and 296 first confirmations}\label{sec:headline}

We call an object \emph{new} if it has no previous brown-dwarf identification or candidacy in SIMBAD \citep{Wenger2000}, the UltracoolSheet \citep{UltracoolSheet}, the 20-pc census \citep{Kirkpatrick2021,Kirkpatrick2024}, SPIFF \citep{Gagne2026}, Backyard Worlds \citep{Kuchner2017,Schneider2026}, the candidate lists of \citet{Griffith2012}, \citet{CarneroRosell2019}, \citet{DalPonte2023}, \citet{Brooks2024} and \citet{Karpov2025}, in any VizieR catalogue \citep{Ochsenbein2000} of brown dwarfs, or among the designations in the arXiv sources of brown-dwarf papers. Objects previously published only as photometric or motion candidates, without a spectrum, are \emph{first spectroscopic confirmations}; 108 of them are Backyard Worlds discoveries of \citet{Schneider2026}, motion-confirmed candidates whose types were estimated from their colours, and 58 are dwarfs whose published types are photometric only (Sect.~\ref{sec:val_known}). Objects with a published spectral type, including those typed from SPHEREx spectra in SPIFF, are known (Sect.~\ref{sec:val_known}). None of our new objects or confirmed candidates is in SPIFF \citep{Gagne2026}.

\begin{figure}[!htbp]
\centering
\includegraphics[width=\columnwidth]{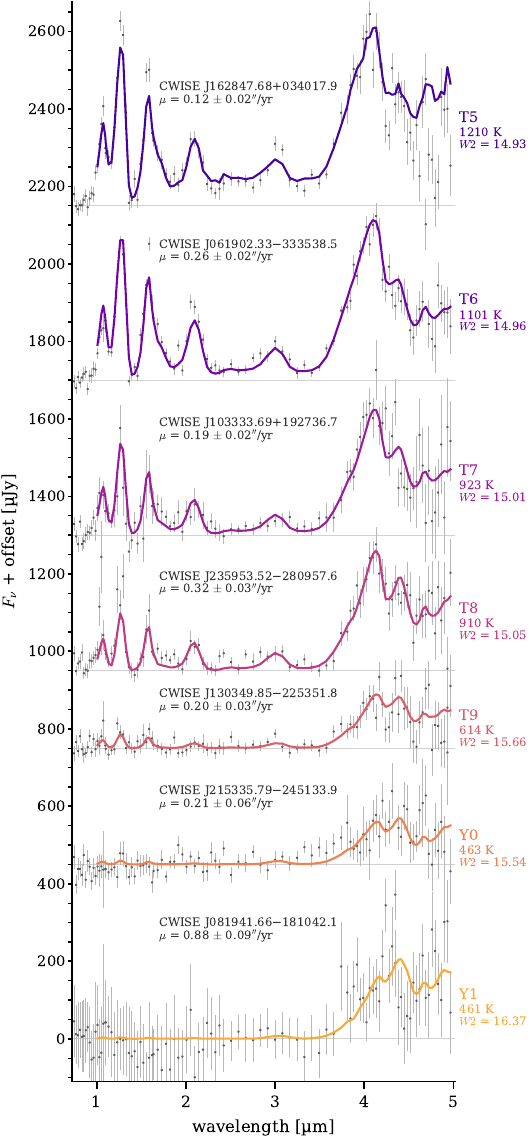}
\caption{SPHEREx spectra of seven of the new Detections, one of each type from T5 to Y1 (among the highest-S/N Detections of each type, with a significant motion and a clean spectrum), from the hottest (top) to the coldest: measured spectrum (grey points, with 1$\sigma$ errors) and best template scaled to it (coloured by type, as in Fig.~\ref{fig:stacks}). All spectra are on the same flux scale and shifted upwards by round offsets; the thin grey line under each spectrum marks its zero level, so its flux is read as the height above that line. Their peak fluxes therefore follow $W2$ and type: the colder the dwarf, the larger the fraction of its flux in the 4--5\,\micron\ peak. Left labels: designation and the motion between CatWISE2020 and SPHEREx; right labels: template, Elf Owl temperature and CatWISE2020 $W2$.}
\label{fig:spectra}
\end{figure}

\begin{figure*}[!tbp]
\centering
\includegraphics[width=\textwidth]{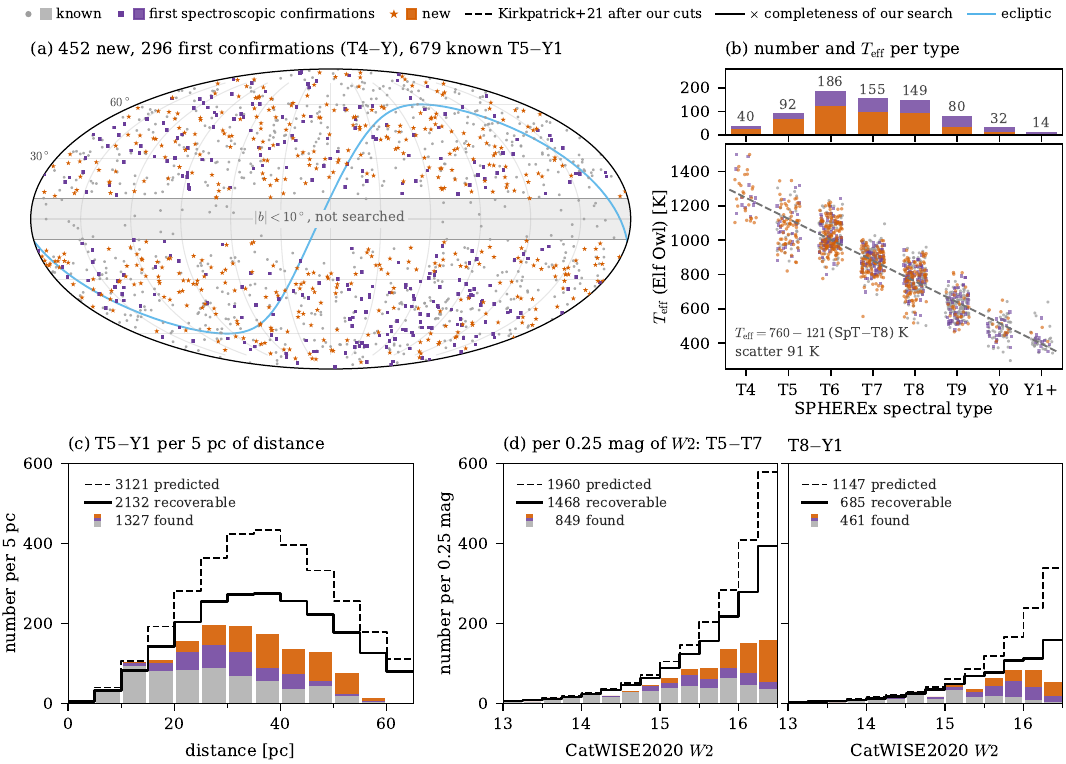}
\caption{What the search found. (a) Sky distribution in Galactic coordinates (Mollweide projection, $l=0^\circ$ at the centre, $l$ increasing to the left) of the new T and Y dwarfs (orange stars), the first spectroscopic confirmations (purple squares) and the known T5--Y1 dwarfs of our merged list (grey dots), all with $W2<16.5$. (b) Top: number of new dwarfs (orange) and first confirmations (purple) per SPHEREx type; the last bin includes the two confirmations typed later than Y1 (one Y4 and one Y pec). Bottom: their Elf Owl effective temperature against type, with the known T and Y dwarfs typed in the same way (grey); dashed line: a robust linear fit to our T5--Y1 Detections, $T_{\rm eff}=760-121\,({\rm SpT}-{\rm T8})$\,K, with a scatter of 91\,K. (c), (d) The census of T5--Y1 dwarfs with $W2<16.5$ at $|b|>10^\circ$ (Sect.~\ref{sec:census}): (c) per 5\,pc of distance (parallax distances for the known dwarfs when better than 20\%, otherwise spectrophotometric) and (d) per 0.25\,mag of CatWISE2020 $W2$ (from $W2=13$), for T5--T7 (left) and T8--Y1 (right; the T4 dwarfs are not included), all on a common vertical scale. Stacked bars: known dwarfs that pass the colour cut (grey), first confirmations (purple) and new dwarfs (orange), Detections only. Dashed lines: the numbers predicted after our cuts by the space densities of \citet{Kirkpatrick2021}; solid lines: the same times our completeness, i.e.\ what our search should find; the key in each panel gives these totals and the number found. Panel (d) omits the 19 known dwarfs brighter than $W2=13$, and panel (c) the two dwarfs fitted best by the Y4 template, which have no distance; the expectations in (c), integrated over distance out to 65\,pc, differ slightly from those of Table~\ref{tab:census}.}
\label{fig:results}
\end{figure*}

The search finds 452 new T and Y dwarfs and confirms spectroscopically, for the first time, 296 photometric candidates (Table~\ref{tab:headline}); another 228 objects are tentative (Table~\ref{tab:fulltent}). Figure~\ref{fig:spectra} shows examples: their spectra show the features of their types. Of the new objects 27, and of the confirmations 13, are T4 dwarfs; since the colour selection targets T5 and later types, all comparisons and statistics below use only the 425 new and 283 confirmed T5--Y dwarfs. All objects are listed in Appendix~\ref{app:tables}; the tentative ones are not used in any statistic either. Two former photometric candidates, CWISE~J140118.30+432553.5 and J120444.32$-$235927.3, already have near-infrared spectra and spectral types in \citet{Burgasser2025} (d/sdT5.5 and d/sdT7), so we count them as known. Of the new T5--Y dwarfs, 256 have $W2=16$--16.5, as do 93 of the confirmed T5--Y candidates. Figure~\ref{fig:results} compares them with the known dwarfs: they cover the high-latitude sky like the known late-T dwarfs (a), follow the same temperature sequence (b), and, in every spectral class, the known ones peak at $W2\approx15$ and the new ones and the confirmations at $W2=15.5$--16.5, where motion surveys are least complete (d).

\begin{table}[!htbp]
\caption{New T and Y dwarfs and first spectroscopic confirmations ($W2<16.5$, $|b|>10^\circ$).}
\label{tab:headline}
\centering\small\setlength{\tabcolsep}{5pt}
\begin{tabular}{lcccc}
\toprule
 & \multicolumn{2}{c}{New} & \multicolumn{2}{c}{Confirmed} \\
\cmidrule(lr){2-3}\cmidrule(lr){4-5}
Class & Det. & Tent. & Det. & Tent. \\
\midrule
T4 & 27 & 12 & 13 & 0 \\
T5--T6 & 189 & 34 & 89 & 8 \\
T7--T8 & 189 & 43 & 115 & 5 \\
T9 & 32 & 35 & 48 & 8 \\
Y & 15 & 72 & 31 & 11 \\
\midrule
Total & 452 & 196 & 296 & 32 \\
\quad of which T5--Y$^{a}$ & 425 & 184 & 283 & 32 \\
\bottomrule
\end{tabular}
\tablefoot{New: no earlier brown-dwarf identification or candidacy; Confirmed: published photometric or motion candidates, for which our SPHEREx spectra are the first spectra; dwarfs with a published spectroscopic type are known and not counted here. Det.: Detections; Tent.: tentative objects; class of the best-fitting template (SPHEREx types, which run late with respect to the near-infrared scale; Sect.~\ref{sec:val_known}), except two Y-type objects detected in $J$ that are counted as T9 from their $J-W2$ (Sect.~\ref{sec:props}). $^{a}$The types targeted by the colour selection; only these enter the comparisons and statistics.}
\end{table}

\subsection{Their properties}\label{sec:props}

The new objects are faint, $W2=14.7$--16.5 (median 16.1): earlier surveys had already found the bright ones (Sect.~\ref{sec:census}). Fits of the 275--1500\,K part of the Sonora Elf Owl grid \citep{Mukherjee2024} place them along the same temperature sequence as the known dwarfs (Fig.~\ref{fig:results}b), with medians of 1040\,K for the T5--T6, 820\,K for the T7--T8, 610\,K for the T9 and 490\,K for the Y-type dwarfs; those typed T5--Y by SPHEREx span 400--1350\,K. These temperatures are approximate and model-dependent: they come from one model grid fitted to low-resolution spectra, and for late-T and Y dwarfs the absolute $W2$ magnitude is a more robust temperature indicator \citep{Leggett2025}; we use them to place the objects on the temperature sequence, not as precise measurements. Their photometric distances, from the absolute magnitude of each SPHEREx type calibrated on the known dwarfs with parallaxes (Sect.~\ref{sec:val_known}; scatter 19\%, Fig.~\ref{fig:validation}e), are 14--57\,pc (median 39\,pc); 34 lie within 25\,pc; the nearest of them are of Y type (see below). The confirmed candidates are similar (median 31\,pc; 72 within 25\,pc).

We also looked for our objects in the first data release of Euclid \citep{Mellier2025,EuclidDR1}: the 56 Detections with a Euclid counterpart all have the colours expected for cold T dwarfs, those of the known T dwarfs in the same data ($\mathrm{VIS}-J=4.3$--5.5 and $J-H\leq0.2$, AB), whereas the three rejected candidates that Euclid covers are bluer, like galaxies. An independent search of the Euclid DR1 fields for ultracool dwarfs, also carried out at our institute \citep{DominguezTagle2026}, which reaches fainter magnitudes and earlier types, shares 13 of our new dwarfs and 10 of our confirmations, and all its T5 and later discoveries within our selection are also Detections here, except one blended with an unrelated star 1\arcsec\ away, which we list as Tentative. In a concurrent study, \citet{ZhengGu2026} identified 18 of our Detections (9 new and 9 confirmations, all T4--T7) as T dwarfs in the Euclid DR1 slitless spectra; their types agree with ours within one subtype for 13 of them, ours being 0.9 subtypes later on average, as against the literature types (Sect.~\ref{sec:val_known}).

Twenty new dwarfs move faster than 80\,km\,s$^{-1}$ (motions significant at $\geq3\sigma$), velocities more common in the thick disk; they are candidates for old, metal-poor dwarfs, a class with very few known members \citep{Burningham2010,Scholz2010,Pinfield2012,Mace2013,Schneider2020,Meisner2020,Meisner2021,Brooks2022,Lodieu2022,Burgasser2025}, although photometric distances and motions selected at $\geq3\sigma$ favour upward errors. The fastest is the T9 dwarf CWISE~J203224.66$-$243224.5 ($\approx$130\,km\,s$^{-1}$; 0.86\arcsec\,yr$^{-1}$ at 9$\sigma$); the most significant motions are those of CWISE~J163857.95$-$074210.2 (T8, $\approx$112\,km\,s$^{-1}$, 22$\sigma$), CWISE~J123654.44$-$202808.9 (T8, $\approx$106\,km\,s$^{-1}$, 18$\sigma$) and CWISE~J064130.10+675919.8 (T6, $\approx$104\,km\,s$^{-1}$, 15$\sigma$), and the Y0-type dwarf CWISE~J123543.50$-$162524.6 moves at $1.07\pm0.09$\arcsec\,yr$^{-1}$ ($\approx$100\,km\,s$^{-1}$ at 20\,pc). None of them shows the suppressed $K$ band of T subdwarfs at the SPHEREx resolution, but our templates are of solar metallicity, so their types need care \citep{Burgasser2025}. The halo brown dwarf WISEA~J153429.75$-$104303.3 (`The Accident'; $[\mathrm{M/H}]\approx-2.2$, $T_\mathrm{eff}\approx500$\,K, $v_\mathrm{tan}\approx210$\,km\,s$^{-1}$; \citealt{Kirkpatrick2021b,Faherty2025}) lies within our selection ($W2=16.15$) but is detected only marginally (template S/N 6.2). Fitting its JWST spectrum, at SPHEREx resolution, to every detected target that moves at $\geq5\sigma$ finds no new analogue (although a search with solar-metallicity templates is itself less sensitive to such spectra): it fits better than our templates only the two known metal-poor dwarfs in our footprint, Wolf~1130C (sdT8, \citealt{Mace2013}; (e)sdT6: in \citealt{Burgasser2025}) and WISEA~J041451.67$-$585456.7 (esdT0, \citealt{Schneider2020}; esdT6: in \citealt{Burgasser2025}).

The Y-type objects need particular care, because at low S/N a red continuum rising to 5\,\micron\ can resemble a Y-dwarf spectrum. Besides requiring a motion confirmed by eye, we checked every T9- and Y-type object against Pan-STARRS1 \citep{Chambers2016}, the Legacy Surveys DR10 $z$ band \citep{Dey2019}, AllWISE $W3$ \citep{Cutri2013} and Gaia DR3 (Appendix~\ref{app:score}): a $z$ counterpart, a red $W2-W3$ or a Gaia source at the position rules out a Y dwarf this faint. Contaminants that reach the inspection end up preferentially with T9 and Y templates: all 11 blank positions that pass the vetting do (Sect.~\ref{sec:val_inj}). This is why a Y-type Detection requires a motion confirmed by eye, and why we expect a good fraction of the Tentative Y-type objects to be contaminants. Seventeen new Detections are best fitted by Y0--Y1 templates, all with a motion significant at $\geq3\sigma$. Four of them are detected in $J$. Two are bluer than almost all known Y dwarfs with $J$ photometry (the 23 known Y0 dwarfs have $J-W2=3.3$--8.1, median 5.1): CWISE~J231239.38+103205.9 ($J-W2=2.5$) is bluer than all of them and than every known Y1 dwarf, and J142846.68$-$103443.7 (3.4) than all but one Y0 dwarf. We list both as T9, without a temperature, since the model fit to their spectra is Y-like; the other two ($J-W2=4.0$ and 4.1) lie within the Y0 range but also near the T9 median (4.1), and may be either. Of the 15 candidate Y dwarfs, the type calibration of Sect.~\ref{sec:val_known} predicts that about six are Y dwarfs and the others T8--T9 dwarfs. The confirmed candidate CWISE~J001146.08$-$471306.6 clearly moves ($284\pm46$\,mas\,yr$^{-1}$, $6.2\sigma$) but no single template fits its spectrum (best Y4, with Y1 and T9 almost as good), nor does the JWST spectrum of The Accident, despite a similarly steep rise to 4\,\micron, so we list it as a peculiar Y-type dwarf (Y pec) without a spectrophotometric distance. Two of the nearest, CWISE~J235933.51+172115.3 (Y1, $\approx$490\,K, 14\,pc) and CWISE~J054417.59+003623.4 (Y1, $\approx$410\,K, 16\,pc), move only about 0.2\arcsec\,yr$^{-1}$, slow enough for motion surveys to miss them easily, and are targets for immediate follow-up. The Tentative Y-type objects without motion are the least secure: little or no flux below 3.5\,\micron\ and a steep rise to 4\,\micron\ could also come from very red extragalactic sources or heavily reddened young stars, although these are usually bright at 12\,\micron. All the new objects are identified by low-resolution spectra and, for 54\% of the new T5--Y dwarfs, also by their motion; follow-up spectroscopy will refine their types and remove any remaining contaminants.

\subsection{The SPHEREx spectral sequence}\label{sec:stacks}

\begin{figure}[!htbp]
\centering
\includegraphics[width=\columnwidth]{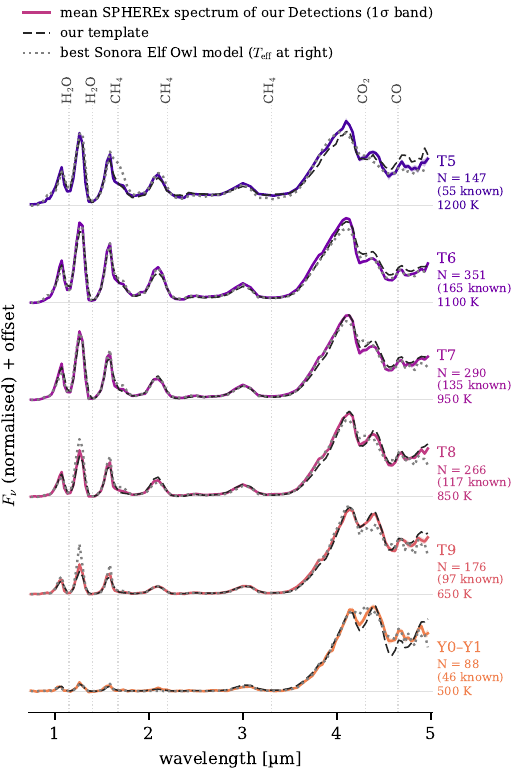}
\caption{The SPHEREx spectral sequence T5--Y1 defined by the data: the mean SPHEREx spectrum of all our Detections of each SPHEREx type, with Y0 and Y1 averaged together (new dwarfs, first confirmations and recovered known dwarfs; coloured, with its bootstrap 1$\sigma$ band; N: number of dwarfs, and how many of them are known), our template of the same type (Y0 for the Y0--Y1 group; dark dashed) and the best Sonora Elf Owl model (grey dotted; its temperature is given under N). Each spectrum is scaled to the mean of its type before averaging, and each mean is normalised to its peak and offset, so the vertical axis has no physical flux scale (the absolute fluxes of individual objects are shown in Fig.~\ref{fig:spectra}); the main molecular bands are marked at the top, as in Fig.~\ref{fig:templates}. The types are SPHEREx types, which run late at T9--Y (Sect.~\ref{sec:val_known}): the Y0--Y1 group contains many dwarfs typed T8 or T9 in the literature, so it is not a sample of Y dwarfs.}
\label{fig:stacks}
\end{figure}

Averaging the spectra of all our Detections of each type (the new dwarfs, the first spectroscopic confirmations and the known dwarfs that we detect, each counted once) gives a sequence of mean SPHEREx spectra defined by the data (Fig.~\ref{fig:stacks}). We grouped the dwarfs by their SPHEREx type, the only type available for all of them, and averaged those with a template S/N of at least 5, from 88 (Y0--Y1) to 351 (T6) per group, each scaled to the mean of its type, with inverse-variance weights and 3$\sigma$ clipping; the errors come from bootstrap resampling of the objects. We average Y0 and Y1 together, since at our S/N they cannot be told apart reliably. This group is defined by our fit, and it holds late-T dwarfs as well as Y dwarfs: half of its 46 known members (23) are T8 or T9 in the literature, so its size (88, with 42 new or confirmed dwarfs, 14 of them new) is not a count of Y dwarfs. Because the dwarfs are grouped by their best template, the averages are not independent of our templates; grouping the known dwarfs by their literature types instead matches the templates less well for T5 and T8. The averages are released with the spectra.

Our templates reproduce the averages to a reduced $\chi^2$ of 1.6--2.6 (with a 10\% allowance for the template uncertainty); the largest deviations are an excess at 3.6--4.1\,\micron\ and, for T6 and T7, a deficit at 4.2--4.4\,\micron\ (Appendix~\ref{app:excess}). Nor do the templates reproduce in detail the shape of the $J$ and $H$ peaks (Figs.~\ref{fig:spectra} and \ref{fig:stacks}), so empirical templates built from these averages could make a future search somewhat more sensitive. Our templates, like the Elf Owl models we fit, are of solar metallicity; low-metallicity or low-gravity templates (for example Sonora models with $[\mathrm{M/H}]=-1$, or the JWST spectrum of Wolf~1130C) may account for part of these deviations. The best Sonora Elf Owl models \citep{Mukherjee2024} follow the sequence from 1200\,K at T5 to 500\,K at Y0--Y1 and fit the averages from T6 to Y1 about as well as or better than our templates, but not that of T5.

The new dwarfs are spectroscopically the same population as the known ones: averaged in the same way, the new Detections of each type from T5 to T9 agree with the averages of the known dwarfs of that type to a reduced $\chi^2$ of 0.8--2.7.

Many T6--T8 dwarfs, known and new alike, are brighter than their best template between 3.6 and 4.1\,\micron, typically by 4--11\% of the peak flux, and fainter just beyond the peak (Appendix~\ref{app:excess}); the mean spectra of Fig.~\ref{fig:stacks} show the same pattern. The excess is as clear in the known dwarfs as in the new ones, so it is not a property of our selection, and we find no instrumental origin: in the individual exposures it rises smoothly across the boundary between detectors 4 and 5 at 3.8\,\micron, where both give the same excess, and it is the same in the first and second half of the observations. The templates of \citet{Gagne2026} thus do not describe the average T6--T8 dwarf in this range: typical T6--T8 dwarfs have less absorption in the red wing of the 3.3\,\micron\ methane band, as expected if disequilibrium chemistry, gravity or metallicity vary across the population. We leave its interpretation to a dedicated study.

\section{What the new objects tell us}\label{sec:science}

\begin{table*}[!tbp]
\caption{The census of T5--Y1 dwarfs with $W2<16.5$ at $|b|>10^\circ$.}
\label{tab:census}
\centering\small
\sbox0{\begin{tabular}{lccccccc}
\toprule
Class & Known$^{a}$ & Added$^{b}$ & Found & Expected$^{c}$ & Found / & Expected $\times$ & Found / (Expected \\
 & & & & & Expected & completeness$^{d}$ & $\times$ completeness) \\
\midrule
T5--T6 & 319 & 278 & 597 & 1409 & 42\% & 1102 & 54\% \\
T7--T8 & 227 & 304 & 531 & 1343 & 40\% & 893 & 59\% \\
T9--Y1 & 75 & 126 & 201 & 355 & 57\% & 159 & 126\% \\
\midrule
Total & 621 & 708 & 1329 & 3107 & 43\% & 2154 & 62\% \\
\bottomrule
\end{tabular}}\usebox0\par\vspace{2pt}
\parbox{\wd0}{\scriptsize $^{a}$Known T5--Y1 dwarfs (with a published spectroscopic type; Sect.~\ref{sec:val_known}) that pass our colour cut in the photometry of our merged list (or have no $W1$), and the four dwarfs added in Sect.~\ref{sec:census}; they are more than the 605 of Table~\ref{tab:val}, which meet our full selection in CatWISE2020. $^{b}$New objects and confirmed candidates not in the known sample (Detections), classified by their SPHEREx types, which run late (Sect.~\ref{sec:census}); the known dwarfs are classified by their literature types. $^{c}$Space densities and absolute magnitudes per subclass of \citet{Kirkpatrick2021} after our cuts in magnitude, colour and latitude (uncertainty from the densities 10--13\% per class, 7\% in total); beyond $W2=16$ extended with the $10^{0.6\,\Delta m}$ slope of a uniform density. $^{d}$Multiplied by the completeness of the search measured with the injected spectra (Sect.~\ref{sec:val_inj}).}
\end{table*}

\subsection{The gap left by motion selection}\label{sec:kinematics}

Of the 425 new T5--Y dwarfs, 230 move at $\geq3\sigma$ in our motion fit (CatWISE2020, SPHEREx and, where available, $J$-band positions); for the other 195 our precision (median quoted uncertainty per component 39\,mas\,yr$^{-1}$ at $W2=15.5$--16 and 61\,mas\,yr$^{-1}$ at 16--16.5) does not detect a motion. The known T dwarfs were found mostly by their motion, so these could be the slow dwarfs that motion surveys miss. If a fraction $f$ of the T dwarfs in a distance shell moves too slowly to pass a $5\sigma$ CatWISE2020 motion cut, the known ones are a fraction $1-f$ of all, and we should find $c\,f/(1-f)$ times as many new ones, where $c$ is our completeness. We measure $f$ by assuming that field T dwarfs move like the Gaia DR3 M dwarfs within 50\,pc at $|b|>10^\circ$, a far more complete thin-disk sample with a median tangential velocity (32\,km\,s$^{-1}$) close to that of the known T5--Y dwarfs (33\,km\,s$^{-1}$). With the CatWISE2020 motion precision at their $W2$, the fraction missed rises from 3\% at 10--15\,pc to 85\% at 40--45\,pc.

The new objects fill much of this gap at 20--30\,pc: we find 21 where $32\pm3$ are expected at 20--25\,pc and 46 where $82\pm7$ are expected at 25--30\,pc, and about a third of the expected numbers at 30--40\,pc and a sixth at 40--50\,pc. Closer in, few dwarfs are missing at all, because the larger motions of nearby dwarfs are found by searches more sensitive than the catalogue cut we assume. Their velocities agree: the new T dwarfs have a median tangential velocity of 30\,km\,s$^{-1}$ (all measured motions, significant or not), close to the 26\,km\,s$^{-1}$ of the simulated dwarfs that a motion cut misses and below the 32\,km\,s$^{-1}$ of the whole population, and a Kolmogorov--Smirnov test does not distinguish the two distributions ($p=0.55$). The new objects are thus consistent with being the slow T dwarfs that motion surveys have missed. Beyond 25\,pc, known and new dwarfs together still fall short of the expected numbers by more than the slow dwarfs explain: the low S/N of both the WISE and the SPHEREx data at $W2=15$--16.5 limits the census there (Sect.~\ref{sec:census}).

\subsection{How complete is the census now?}\label{sec:census}

Our search increases the number of known T5--Y1 dwarfs with $W2<16.5$ at $|b|>10^\circ$ by 104\%, from 679 to 1387: it adds the 425 new and the 283 confirmed T5--Y dwarfs. The 679 known dwarfs are the known sample of Sect.~\ref{sec:val_known} (675), whether or not our search recovers them, plus four dwarfs with published spectroscopic types that are in neither list: CWISE~J140118.30+432553.5 and J120444.32$-$235927.3 (Sect.~\ref{sec:headline}), CWISE~J225109.59$-$074037.2 (d/sdT7; \citealt{Burgasser2025}) and ULAS~J080048.27+190823.8 (T6.5; \citealt{Burningham2013}); the known dwarfs of Fig.~\ref{fig:flow} are only those among our Detections.

Known and new dwarfs together, 1329 T5--Y1 dwarfs that pass our colour cut, are 62\% of the number that a search with our completeness should find, and 43\% of the number predicted after our cuts (Table~\ref{tab:census}; Fig.~\ref{fig:results}c,d). Both counts and expectations refer to our colour cut ($W1-W2\geq1.5$), which leaves out a significant fraction of the T5 and T6 dwarfs (Sect.~\ref{sec:val_known}). Both expectations come from the space densities of \citet{Kirkpatrick2021}, with the absolute magnitude of each subclass and our cuts in magnitude, colour and latitude; the first is multiplied by our completeness measured with the injected spectra (Sect.~\ref{sec:val_inj}). At the bright end ($W2<14.6$), where our search is about 85\% complete, the census is 80\% of the expectation. The expectation is sensitive to the adopted absolute magnitudes: a magnitude-limited count scales as $10^{0.6\,\Delta M}$, so a shift of only 0.2\,mag in $M_{W2}$ changes it by 30\%, more than the uncertainty from the densities (10--13\% per class, 7\% in total).

Within 15\,pc the known dwarfs match the expected number (Fig.~\ref{fig:results}c), as they must, since the space densities come from the 20-pc census itself. Beyond 15\,pc it falls behind, and the missing dwarfs are at $W2=15$--16, the faint end of both kinds of search, where a motion of 100\,mas\,yr$^{-1}$ is only a $2\sigma$ measurement in CatWISE2020 and our completeness falls to 50--80\%. Most of our losses there are due to the low S/N; confusion with brighter neighbours accounts for a minority of them (a brighter source lies within one SPHEREx pixel of 4--7\% of the dwarfs at $W2=15.5$--16, against injection losses of 22--54\%). The search also inherits the limits of CatWISE2020, on which both the selection and the measurement rest: a dwarf that is missing from the catalogue, blended with a neighbour in it, fragmented (Sect.~\ref{sec:measurement}) or given a wrong colour cannot be found.

By class, the census reaches 54\% (T5--T6), 59\% (T7--T8) and 126\% (T9--Y1) of what our search should find ($\pm6$, $\pm6$ and $\pm16$ percentage points from the density errors alone). For T9--Y1 the census thus exceeds the expected number, by about a quarter (1.6$\sigma$ from the density errors alone), although the coldest dwarfs are the hardest to detect. This excess most likely comes from our typing. SPHEREx types run late (Sect.~\ref{sec:val_known}), so some late-T dwarfs among the new and confirmed ones are counted as T9--Y1, whereas the known dwarfs are counted by their literature types: binned by their SPHEREx types, the known dwarfs that we detect would number 130 in T9--Y1 instead of 68. Scaling the added dwarfs of each class by this ratio brings the census to about 67\%, 56\% and 89\% for the three classes, below the expectation for T9--Y1 as for the warmer types, so the typing alone accounts for the excess. Three other effects may contribute. The expected number may be too low: \citet{Kirkpatrick2021} measured the T8 and T9 densities within 16.5 and 17.5\,pc, and the Y0 and Y1 densities within only 12 and 13.5\,pc and from only 16 and 11 objects, and the 20-pc census of the coldest dwarfs may itself be incomplete. Some of our Y-type Detections may be contaminants, despite the checks of Sect.~\ref{sec:props}. Or the mass function may produce more cold dwarfs than assumed. Spectroscopy and parallaxes of the new Y-type dwarfs will tell these apart.

The coming SPHEREx data will recover many of the missing dwarfs. The nominal two-year mission ends with a fourth all-sky pass, which raises the S/N of a source by about $\sqrt{4/3}\approx1.15$, and any extended mission adds more: with six passes our completeness limit would move about 0.4\,mag fainter, enough to reach at $W2=15.5$--16 the completeness that we now reach at $W2=15$--15.5. A volume-limited analysis per type, with the scale height of the disk and the colour scatter as free parameters, is left to future work.

\subsection{Wide companions}\label{sec:wide}
\begin{table*}[!tbp]
\caption{Candidate wide stellar companions of the brown dwarfs (Sect.~\ref{sec:wide}).}\label{tab:wide}
\centering\small\setlength{\tabcolsep}{1.7pt}
\sbox0{\begin{tabular}{lcccr@{\hspace{2pt}}l@{\hspace{2pt}}lcccr@{\hspace{2pt}}lc}
\toprule
Brown dwarf & SpT$^{a}$ & $W2$ & $d^{b}$ & \multicolumn{2}{c}{$\mu_{\alpha}^{*}$, $\mu_{\delta}$} & Gaia DR3 star & $G$ & $G_{\rm BP}{-}G_{\rm RP}$ & $\varpi$ & \multicolumn{2}{c}{$\mu_{\alpha}^{*}$, $\mu_{\delta}$} & Separation \\
 & & (mag) & (pc) & \multicolumn{2}{c}{(mas\,yr$^{-1}$)} & (source id) & (mag) & (mag) & (mas) & \multicolumn{2}{c}{(mas\,yr$^{-1}$)} & ($''$ / au) \\
\midrule
WISE J235447.80$-$814044.9$^{(1)}$ & T8 & 15.03 & 20.5 & 122, & 71 $\pm$ 26 & 6350786278796634112 & 17.22 & 1.31 & 37.1 & 96, & 36 & 63.9 / 1721 \\
WISEU J001908.31$-$094323.3$^{(2)}$ & T8 & 15.81 & 29.5 & $-$120, & $-$350 $\pm$ 56 & 2428176257763712512 & 15.31 & 3.14 & 22.3 & $-$167, & $-$276 & 18.1 / 814 \\
CWISE J104053.42$-$355029.7$^{(3)}$ & T5 & 14.94 & 24.0 & $-$135, & $-$6 $\pm$ 10 & 5443160355149610368 & 13.23 & 2.87 & 39.6 & $-$129, & 11 & 65.6 / 1659 \\
\midrule
CWISE J004058.63$-$294232.0$^{(4)}$ & T6 & 15.91 & 41.3 & 126, & $-$90 $\pm$ 41 & 2317933586965896832 & 17.96 & 4.29 & 18.2 & 84, & $-$125 & 19.7 / 1086 \\
CWISE J032809.84$-$700025.0$^{(5)}$ & Y1 & 16.15 & 16.3 & 100, & 374 $\pm$ 58 & 4643131608600380800 & 12.00 & 2.41 & 46.1 & 88, & 287 & 71.4 / 1549 \\
CWISE J030651.26+224000.8$^{(6)}$ & T7 & 15.75 & 34.4 & 72, & $-$244 $\pm$ 34 & 109127323626669952 & 13.11 & 2.51 & 26.6 & 15, & $-$259 & 110.2 / 4143 \\
CWISE J015458.74+301634.0$^{(7)}$ & T6 & 16.18 & 46.7 & 232, & 24 $\pm$ 69 & 301022369307171840 & 9.15 & 1.03 & 17.6 & 185, & $-$10 & 80.8 / 4590 \\
\bottomrule
\end{tabular}}\usebox0\par\vspace{2pt}
\parbox{\wd0}{\raggedright\scriptsize WISE~J2354$-$8140 and CWISE~J1040$-$3550 are known brown dwarfs; the other five are new or confirmed here. Stars and literature: (1) a cool DZH white dwarf \citep{OBrien2023}, new candidate pair; (2) LP~704-85, common motion noted by \citet{Meisner2020}, candidate pair (parallaxes agree within $1.7\sigma$); (3) UPM~J1040$-$3551, known pair \citep{Rothermich2024}; (4) LEHPM~877, known pair \citep{Kiwy2022}; (5) L~55-11 (M3V), known pair \citep{Marocco2024}; (6) LP~355-25, new candidate pair; (7) TYC~2307-994-1, new candidate pair. Motions: ours for the new and confirmed dwarfs, the literature values for the known ones; Gaia DR3 values for the stars. $^{a}$SPHEREx type. $^{b}$Spectrophotometric distance; the two known dwarfs have no parallax.}
\end{table*}

Wide companions of stars make brown dwarfs into benchmarks, and how often they occur constrains how brown dwarfs form. Around the 990 brown dwarfs of our catalogue that move faster than 100\,mas\,yr$^{-1}$ (232 new and 249 confirmed, T4 dwarfs included, with motions significant at $\geq3\sigma$, and 509 known) we searched Gaia DR3 from 2 to 1000\arcsec\ for stars whose proper motion agrees with that of the dwarf within $2\sigma$ and whose parallax agrees with its distance. The same search in control fields 1.5$^\circ$ away, where every pair is a chance alignment, gives the number of chance pairs expected at each separation.

For the 293 known brown dwarfs with parallaxes the search finds 11 pairs where one chance pair is expected, all of them known benchmark companions, from Wolf~940B to Ross~458C and COCONUTS-2b. For most of them, however, the literature motion and parallax of the dwarf are those of its primary, so their recovery tests the search rather than our astrometry; our own SPHEREx motions agree with those of the primaries within $2\sigma$ for seven of them and differ by 2.8--5$\sigma$ for SIPS~J2150$-$7520B and LSPM~J0055+5948B, 15--17\arcsec\ from their primaries, where the stars bias the SPHEREx position, and for SDSS~J1758+4633; Wolf~940B, 32\arcsec\ from its primary, is not among our Detections. The search misses WISE~J1118+3125, whose primary $\xi$~UMa has no Gaia astrometry, and WISEPC~J1122+2550, whose literature motion differs from that of LP~374-39 by 33\,mas\,yr$^{-1}$; our motion of the dwarf, $(\mu_\alpha^*,\mu_\delta)=(-956,-331)\pm54$\,mas\,yr$^{-1}$, agrees with that of the star within about $1\sigma$, so we keep the pair. Companions of bright stars are harder for our method to find, because the blank-position scatter $k$ grows near them: the known T dwarfs LP~381-30~B and LP~270-10~B, about 50\arcsec\ from stars with $G=11$--13, fail the robust threshold. Among the 216 known dwarfs without parallaxes, with our photometric distances, the search finds two more pairs (Table~\ref{tab:wide}). CWISE~J104053.42$-$355029.7 (T5) and the high-proper-motion star UPM~J1040$-$3551 ($G=13.2$), 66\arcsec\ (1660\,au) apart, are a known pair \citep{Rothermich2024}, and the parallax distance of the star (25\,pc) agrees with our photometric one (24\,pc). WISE~J235447.80$-$814044.9 (T8 in SPHEREx), 64\arcsec\ (1700\,au) from a cool DZH white dwarf \citep{OBrien2023} at 27\,pc (our distance of the dwarf: $21\pm4$\,pc), is, to our knowledge, a new candidate pair; few chance pairs are expected (0.004), but the distances agree only within $1.6\sigma$, and a parallax or radial velocity of the dwarf is needed before it can serve as a benchmark.

A star closer than about 7\arcsec, one WISE or SPHEREx resolution element, blends with the dwarf, which sets the inner limit of the search at about 150\,au at the median distance of the known dwarfs (21\,pc); our 1000\arcsec\ window reaches beyond 10\,000\,au for nearly all of them, and all the companions found lie at a few hundred to 9900\,au. We therefore quote the fraction for about 150--10\,000\,au. Counting the 13 known companions above, the known pair of CWISE~J1040$-$3550 and the new candidate of WISE~J2354$-$8140, $2.9^{+1.0}_{-0.7}$\% of the 509 known T5--Y dwarfs with distances have a stellar companion in this range, and $4.1^{+1.7}_{-1.3}$\% of those within 20\,pc (68\% Clopper--Pearson intervals). Both are lower limits, because companions of bright stars are partly missed. They are below the minimum of 8\% that \citet{Burningham2013} derived for UKIDSS T dwarfs out to 20\,000\,au, and we detect no excess for T8 and later types, which in their sample rested on three objects: the fractions for T5--T8 and T9--Y1 dwarfs (9 of 367 and 6 of 142) agree within their errors. For comparison, 2--3\% of stars have a wide substellar companion \citep{Radigan2008,MetchevHillenbrand2009}.

For the new and confirmed brown dwarfs, whose motions and photometric distances are less precise, the wide pairs are as many as expected by chance, and only five close pairs stand out (Table~\ref{tab:wide}; 0.06, 0.004, 0.13, 0.004 and 0.07 chance pairs expected at their separations, in the order discussed here). CWISE~J032809.84$-$700025.0 (Y1 in SPHEREx), 71\arcsec\ (1550\,au) from the M3V star L~55-11 ($G=12.0$), is a known pair \citep{Marocco2024}, which our search recovers blindly; the parallax distance of the star (21.7\,pc) exceeds ours of the dwarf ($16\pm3$\,pc). CWISE~J004058.63$-$294232.0 (T6), 20\arcsec\ (1100\,au) from the red star LEHPM~877 ($G=18.0$, 55\,pc, against our $41\pm8$\,pc), is also a known pair \citep{Kiwy2022}. The third, to our knowledge, is a new candidate pair: CWISE~J030651.26+224000.8 (T7), 110\arcsec\ (4100\,au) from LP~355-25 ($G=13.1$, 37.6\,pc, against our $34\pm7$\,pc), which lies towards the Hyades but is not a member (it moves 77$^\circ$ away from the convergent point of the cluster, and its Gaia radial velocity is 13\,km\,s$^{-1}$ below the Hyades value). It needs a radial velocity or a parallax of the brown dwarf for confirmation. The fourth is WISEU~J001908.31$-$094323.3 (T8), a photometric candidate of \citet{Meisner2020} that our spectrum confirms as a brown dwarf, 18\arcsec\ (810\,au) from LP~704-85 ($G=15.3$, 45\,pc, against our $30\pm6$\,pc); \citet{Meisner2020} noted the common motion without being able to confirm the association, and the pair passes our criteria, but the parallaxes agree only within $1.7\sigma$, so the association too remains a candidate. The fifth is the new T6 dwarf CWISE~J015458.74+301634.0, 81\arcsec\ (4600\,au) from the $G=9.1$ star TYC~2307-994-1 (57\,pc, against our $47\pm9$\,pc), whose motion agrees with that of the dwarf within $1\sigma$, another new candidate pair. The five pairs are 1.0\% of the 481 new and confirmed dwarfs searched; their true companion fraction is higher, because real pairs at wider separations cannot be told from chance alignments with their less precise motions and distances. We also recover a candidate pair of two brown dwarfs proposed by \citet{Schneider2026}, who could not confirm it because the motion of its fainter component was not significant: CWISE~J052901.09$-$832749.1 (T5 in our SPHEREx fit) and CWISE~J052907.78$-$832740.9, 14\arcsec\ apart. Our SPHEREx spectrum gives the first spectroscopic type of the fainter component, T8, as \citet{Schneider2026} estimated from its photometry, and our motions, $(\mu_\alpha^*,\mu_\delta)=(125,-63)\pm21$ and $(108,-44)\pm35$\,mas\,yr$^{-1}$ ($3.3\sigma$), agree within $0.6\sigma$. With photometric distances of 34 and 32\,pc, the pair would be about 460\,au wide, whereas the known binaries of two T dwarfs are closer than about 15\,au \citep{Schneider2026}; a parallax or radial velocities are needed to confirm it.

At 47\,pc, T5 and later members of the Hyades would lie within our reach, and the search should find some of them. Of the new and confirmed brown dwarfs in the direction of the cluster, three new dwarfs have motions and spectrophotometric distances that both agree with membership: CWISE~J035512.62+262110.4 (T6, 33\,pc), CWISE~J041157.20+222246.6 (T7, 36\,pc) and, only marginally, CWISE~J034112.29+054321.1 (T6, 36\,pc). They lie on the near side of the cluster, as expected for a magnitude-limited sample, and 9--12\% of field brown dwarfs at the same positions, distances and motion uncertainties would pass the same test (simulated thin-disk velocities), so some may be contaminants; radial velocities, predicted to be 34--37\,km\,s$^{-1}$ for members, will decide. If they are members, at the age of the cluster (about 650\,Myr) their types imply masses of about 20--30\,$M_{\rm Jup}$, and they would be the coldest brown dwarf candidates reported in the Hyades: the coldest so far is the T5 candidate CWISE~J042356.23+130414.3 \citep{SchneiderHyades2024}, with $T_{\rm eff}=1136\pm115$\,K and a mass of about 30\,$M_{\rm Jup}$.

\section{Conclusions}\label{sec:conclusions}

We measured our own SPHEREx spectrum, by forced photometry on the Level~2 images, for every red WISE source ($W1-W2\geq1.5$ or no $W1$ detection) to $W2=16.5$ over the whole sky at $|b|>10^\circ$, and searched them for cold brown dwarfs regardless of their motion.
\begin{enumerate}
\item The blind search rediscovers 92\% of the known T5--Y dwarfs that meet its selection (98\% at $W2<15$), with types within one subtype of the literature for 82\% of them and consistent photometry and proper motions; injected spectra calibrate its completeness.
\item It discovers 452 new T and Y dwarfs and gives the first spectra of 296 photometric candidates (425 and 283 of type T5--Y), mostly mid- and late-T dwarfs at 20--55\,pc with temperatures of 400--1350\,K. They increase the number of T5--Y dwarfs known at $|b|>10^\circ$ to $W2=16.5$ by 104\%.
\item Among them are 15 candidate Y dwarfs, the nearest at about 14\,pc (since our fit also types some known late-T dwarfs as Y, we expect about six of them to be Y dwarfs), and 20 dwarfs faster than 80\,km\,s$^{-1}$, candidates for old, metal-poor members of the thick disk; all are prime targets for follow-up spectroscopy.
\item The new dwarfs are the ordinary field population that motion surveys under-sample: their spectra and temperatures follow those of the known dwarfs, and they move more slowly, like the slow dwarfs that a motion cut misses, filling much of the gap that such a cut leaves at 20--30\,pc.
\item Known and new dwarfs together, 1329 T5--Y1 dwarfs, are 62\% of the number that a search with our completeness should find; within 15\,pc the known dwarfs match the expected number, and the dwarfs still missing are faint ($W2=15$--16.5), within reach of further SPHEREx passes.
\item At least $2.9^{+1.0}_{-0.7}$\% of the known T5--Y dwarfs that move faster than 100\,mas\,yr$^{-1}$ have a wide stellar companion in Gaia DR3, at about 150--10\,000\,au, with no excess for the latest types. Two confirmed dwarfs and one new dwarf have candidate stellar companions, two of them not reported before, and a known T dwarf a candidate white-dwarf companion; if confirmed, they would be benchmarks.
\item Our Detections, new and known, define a SPHEREx spectral sequence from T5 to Y1, an empirical reference for typing SPHEREx spectra; SPHEREx types, which include the 3.5--5\,\micron\ region where cold dwarfs emit most of their flux, differ from the literature types by more than one subtype for 18\% of them.
\item The SPHEREx positions give significant proper motions for 475 of the new and confirmed T5--Y Detections, 304 of them without a CatWISE2020 motion.
\item Three new T6--T7 dwarfs have motions and distances consistent with membership of the Hyades; if confirmed, they would be the coldest brown dwarfs known in the cluster, with masses of about 20--30\,$M_{\rm Jup}$.
\end{enumerate}
As prospects: the coming SPHEREx sky passes will deepen the search, and Euclid images can reveal close binaries among the new dwarfs.
With the arXiv version of this paper we release on Zenodo (\href{https://doi.org/10.5281/zenodo.24527431}{doi:10.5281/zenodo.24527431}) the full atlases (a page like Fig.~\ref{fig:example} for every Detection, Tentative object and recovered known dwarf; previewed in Appendix~\ref{app:atlas}), the SPHEREx spectra of all the objects in the atlases in the 102 channels, in both $F_\nu$ and $F_\lambda$, the measurements of Appendix~\ref{app:tables}, a table of the known T5--Y dwarfs recovered by the search (SPHEREx $W2$, type and motion) and the mean SPHEREx spectra of our Detections of each type from T5 to Y1 (Sect.~\ref{sec:stacks}).

\begin{acknowledgements}
This paper would not have been possible without Dr. \'{A}ngel de Vicente, whose key work on the IAC High-Performance Computing infrastructure made a project of this computational scope feasible in the first place; we are grateful to him and to the IAC High-Performance Computing and hardware facilities. We are very grateful to Carlos Dom\'{i}nguez-Tagle for his many technical suggestions and his encouragement. We also thank the participants of the brown-dwarf group meeting at the IAC for their technical comments on this work. We thank the IRSA Help Desk for their assistance with access to the SPHEREx data.
RCA acknowledges funding from the Horizon Europe research and innovation programme under GA 101135036 (RadioForegroundsPlus). RR acknowledges support from the CELESTE project (Cutting Edge Leap to Excellence in Space and Optics Technologies), funded by the European Union as part of the WIDERA Teaming programme under grant agreement 101136736.
This research is based on observations made with the SPHEREx mission and makes use of data products from the SPHEREx Science Data Center, distributed by the NASA/IPAC Infrared Science Archive (IRSA), which is funded by the National Aeronautics and Space Administration and operated by the California Institute of Technology.
This publication makes use of data products from the Wide-field Infrared Survey Explorer, which is a joint project of the University of California, Los Angeles, and the Jet Propulsion Laboratory/California Institute of Technology, funded by the National Aeronautics and Space Administration, and from NEOWISE, which is a joint project of the Jet Propulsion Laboratory/California Institute of Technology and the University of Arizona, funded by the National Aeronautics and Space Administration. We use the CatWISE2020 catalogue and unWISE images.
This work has made use of data from the European Space Agency (ESA) mission Gaia (\url{https://www.cosmos.esa.int/gaia}), processed by the Gaia Data Processing and Analysis Consortium (DPAC, \url{https://www.cosmos.esa.int/web/gaia/dpac/consortium}); funding for the DPAC has been provided by national institutions, in particular the institutions participating in the Gaia Multilateral Agreement. This work has made use of the Euclid DR1 data from the Euclid mission of the European Space Agency (ESA), 2026, \url{https://doi.org/10.57780/esa-d079c51}. It is based in part on observations obtained as part of the VISTA Hemisphere Survey, ESO programme 179.A-2010 (PI: McMahon), and on data obtained as part of the UKIRT Infrared Deep Sky Survey. It uses data from the DESI Legacy Imaging Surveys \citep{Dey2019} and from the Pan-STARRS1 Surveys \citep{Chambers2016}, whose full acknowledgements are given at \url{https://www.legacysurvey.org/acknowledgment} and \url{https://panstarrs.stsci.edu}. This work has benefited from The UltracoolSheet \citep{UltracoolSheet} at \url{http://bit.ly/UltracoolSheet}, maintained by Will Best, Trent Dupuy, Michael Liu, Aniket Sanghi, Rob Siverd, and Zhoujian Zhang, and developed from compilations by \citet{Dupuy2012}, \citet{Dupuy2013}, \citet{Deacon2014}, \citet{Liu2016}, \citet{Best2018}, \citet{Best2021}, \citet{Sanghi2023}, and \citet{SchneiderUHS2023}. We have also used the SIMBAD database and the VizieR catalogue access tool, operated at CDS, Strasbourg, France. We thank the Backyard Worlds: Planet 9 volunteers, whose candidate lists made many of the confirmations possible.
This research makes use of \texttt{astropy} \citep{astropy2013,astropy2018,Astropy2022}, \texttt{numpy} \citep{Harris2020}, \texttt{scipy} \citep{Virtanen2020} and \texttt{matplotlib} \citep{Hunter2007}.
\end{acknowledgements}

\bibliographystyle{aa}
\bibliography{paper_refs}

\begin{thebibliography}{84}
\expandafter\ifx\csname natexlab\endcsname\relax\def\natexlab#1{#1}\fi

\bibitem[{{Akeson} {et~al.}(2026){Akeson}, {Dubois-Felsmann}, {Crill},
  {Faisst}, {et~al.}}]{Akeson2025}
{Akeson}, R., {Dubois-Felsmann}, G.~P., {Crill}, B.~P., {Faisst}, A.~L.,
  {et~al.} 2026, \apjs, 285, 67

\bibitem[{{Assef} {et~al.}(2018){Assef}, {Stern}, {Noirot}, {Jun}, {Cutri}, \&
  {Eisenhardt}}]{Assef2018}
{Assef}, R.~J., {Stern}, D., {Noirot}, G., {et~al.} 2018, \apjs, 234, 23

\bibitem[{{Astropy Collaboration}(2013)}]{astropy2013}
{Astropy Collaboration}. 2013, A\&A, 558, A33

\bibitem[{{Astropy Collaboration}(2018)}]{astropy2018}
{Astropy Collaboration}. 2018, AJ, 156, 123

\bibitem[{{Astropy Collaboration}(2022)}]{Astropy2022}
{Astropy Collaboration}. 2022, ApJ, 935, 167

\bibitem[{{Best} {et~al.}(2025){Best}, {Dupuy}, {Liu}, {Sanghi}, {Siverd}, \&
  {Zhang}}]{UltracoolSheet}
{Best}, W.~M.~J., {Dupuy}, T.~J., {Liu}, M.~C., {et~al.} 2025, The
  {UltracoolSheet}: Photometry, Astrometry, Spectroscopy, and Multiplicity for
  4000+ Ultracool Dwarfs and Imaged Exoplanets, Zenodo, v2.1.0

\bibitem[{{Best} {et~al.}(2021){Best}, {Liu}, {Magnier}, \& {Dupuy}}]{Best2021}
{Best}, W.~M.~J., {Liu}, M.~C., {Magnier}, E.~A., \& {Dupuy}, T.~J. 2021, \aj,
  161, 42

\bibitem[{{Best} {et~al.}(2018)}]{Best2018}
{Best}, W.~M.~J. {et~al.} 2018, ApJS, 234, 1

\bibitem[{{Bock} {et~al.}(2026){Bock}, {Aboobaker}, {Adamo}, {Akeson},
  {et~al.}}]{Bock2025}
{Bock}, J.~J., {Aboobaker}, A.~M., {Adamo}, J., {Akeson}, R., {et~al.} 2026,
  \apj, 999, 139

\bibitem[{{Brooks} {et~al.}(2024){Brooks}, {Caselden}, {Kirkpatrick}, {Raghu},
  {et~al.}}]{Brooks2024}
{Brooks}, H., {Caselden}, D., {Kirkpatrick}, J.~D., {Raghu}, Y., {et~al.} 2024,
  \aj, 168, 211

\bibitem[{{Brooks} {et~al.}(2026){Brooks}, {Cushing}, \&
  {Kothari}}]{BrooksCushingKothari2026}
{Brooks}, H., {Cushing}, M.~C., \& {Kothari}, H. 2026, RNAAS, 10, 94

\bibitem[{{Brooks} {et~al.}(2022){Brooks}, {Kirkpatrick}, {Caselden},
  {Schneider}, {et~al.}}]{Brooks2022}
{Brooks}, H., {Kirkpatrick}, J.~D., {Caselden}, D., {Schneider}, A.~C.,
  {et~al.} 2022, AJ, 163, 47

\bibitem[{{Burgasser} {et~al.}(2025)}]{Burgasser2025}
{Burgasser}, A.~J. {et~al.} 2025, ApJ, 982, 79

\bibitem[{{Burningham} {et~al.}(2013){Burningham}, {Cardoso}, {Smith},
  {et~al.}}]{Burningham2013}
{Burningham}, B., {Cardoso}, C.~V., {Smith}, L., {et~al.} 2013, MNRAS, 433, 457

\bibitem[{{Burningham} {et~al.}(2010){Burningham}, {Leggett}, {Lucas},
  {Pinfield}, {et~al.}}]{Burningham2010}
{Burningham}, B., {Leggett}, S.~K., {Lucas}, P.~W., {Pinfield}, D.~J., {et~al.}
  2010, \mnras, 404, 1952

\bibitem[{{Cardoso} {et~al.}(2015){Cardoso}, {Burningham}, {Smart}, {van
  Spaandonk}, {Baker}, {et~al.}}]{Cardoso2015}
{Cardoso}, C.~V., {Burningham}, B., {Smart}, R.~L., {et~al.} 2015, MNRAS, 450,
  2486

\bibitem[{{Carnero Rosell} {et~al.}(2019){Carnero Rosell}, {Santiago}, {dal
  Ponte}, {Burningham}, {et~al.}}]{CarneroRosell2019}
{Carnero Rosell}, A., {Santiago}, B., {dal Ponte}, M., {Burningham}, B.,
  {et~al.} 2019, \mnras, 489, 5301

\bibitem[{{Chambers} {et~al.}(2016){Chambers}, {Magnier}, {Metcalfe},
  {et~al.}}]{Chambers2016}
{Chambers}, K.~C., {Magnier}, E.~A., {Metcalfe}, N., {et~al.} 2016, The
  {Pan-STARRS1} Surveys, arXiv:1612.05560

\bibitem[{{Cushing} {et~al.}(2011)}]{Cushing2011}
{Cushing}, M.~C. {et~al.} 2011, ApJ, 743, 50

\bibitem[{{Cutri} {et~al.}(2013)}]{Cutri2013}
{Cutri}, R.~M. {et~al.} 2013, Explanatory Supplement to the {AllWISE} Data
  Release Products, IPAC/Caltech

\bibitem[{{dal Ponte} {et~al.}(2023)}]{DalPonte2023}
{dal Ponte}, M. {et~al.} 2023, \mnras, 522, 1951

\bibitem[{{Deacon} {et~al.}(2014)}]{Deacon2014}
{Deacon}, N.~R. {et~al.} 2014, ApJ, 792, 119

\bibitem[{{Dey} {et~al.}(2019)}]{Dey2019}
{Dey}, A. {et~al.} 2019, AJ, 157, 168

\bibitem[{{Dominguez-Tagle} {et~al.}(2026){Dominguez-Tagle}, {Mu{\~n}oz
  Torres}, {Sedighi}, {{\v{Z}}erjal}, {Tsilia}, {Mas-Buitrago}, {Mart{\'\i}n},
  {B{\'e}jar}, {Lodieu}, {Barkaoui}, {Rodr{\'\i}guez-Torres}, \&
  {Burgasser}}]{DominguezTagle2026}
{Dominguez-Tagle}, C., {Mu{\~n}oz Torres}, S., {Sedighi}, N., {et~al.} 2026,
  ApJ, submitted

\bibitem[{{Dupuy} \& {Kraus}(2013)}]{Dupuy2013}
{Dupuy}, T.~J. \& {Kraus}, A.~L. 2013, Science, 341, 1492

\bibitem[{{Dupuy} \& {Liu}(2012)}]{Dupuy2012}
{Dupuy}, T.~J. \& {Liu}, M.~C. 2012, ApJS, 201, 19

\bibitem[{{Dye} {et~al.}(2018){Dye}, {Lawrence}, {Read}, {et~al.}}]{Dye2018}
{Dye}, S., {Lawrence}, A., {Read}, M.~A., {et~al.} 2018, \mnras, 473, 5113

\bibitem[{{Edge} {et~al.}(2013){Edge}, {Sutherland}, {Kuijken},
  {et~al.}}]{Edge2013}
{Edge}, A., {Sutherland}, W., {Kuijken}, K., {et~al.} 2013, The Messenger, 154,
  32

\bibitem[{{Euclid Collaboration} {et~al.}(2026){Euclid Collaboration},
  {Aussel}, {et~al.}}]{EuclidDR1}
{Euclid Collaboration}, {Aussel}, H., {et~al.} 2026, arXiv e-prints
  [\eprint[arXiv]{2611.16510}], {A\&A}, submitted

\bibitem[{{Euclid Collaboration} {et~al.}(2025){Euclid Collaboration},
  {Mellier}, {et~al.}}]{Mellier2025}
{Euclid Collaboration}, {Mellier}, Y., {et~al.} 2025, A\&A, 697, A1

\bibitem[{{Faherty} {et~al.}(2025)}]{Faherty2025}
{Faherty}, J.~K. {et~al.} 2025, Nature, 645, 62

\bibitem[{{Gagn{\'e}} {et~al.}(2026){Gagn{\'e}}, {Faherty}, {Ruiz Diaz},
  {Coulombe}, {et~al.}}]{Gagne2026}
{Gagn{\'e}}, J., {Faherty}, J.~K., {Ruiz Diaz}, A., {Coulombe}, L.-P., {et~al.}
  2026, \apj, 1009, 15

\bibitem[{{Gaia Collaboration} {et~al.}(2016){Gaia Collaboration}, {Prusti},
  {et~al.}}]{GaiaMission2016}
{Gaia Collaboration}, {Prusti}, T., {et~al.} 2016, A\&A, 595, A1

\bibitem[{{Gaia Collaboration} {et~al.}(2023){Gaia Collaboration}, {Vallenari},
  {et~al.}}]{GaiaDR3}
{Gaia Collaboration}, {Vallenari}, A., {et~al.} 2023, A\&A, 674, A1

\bibitem[{{Griffith} {et~al.}(2012){Griffith}, {Kirkpatrick}, {Eisenhardt},
  {Gelino}, {et~al.}}]{Griffith2012}
{Griffith}, R.~L., {Kirkpatrick}, J.~D., {Eisenhardt}, P.~R.~M., {Gelino},
  C.~R., {et~al.} 2012, \aj, 144, 148

\bibitem[{{Harris} {et~al.}(2020)}]{Harris2020}
{Harris}, C.~R. {et~al.} 2020, Nature, 585, 357

\bibitem[{{Hunter}(2007)}]{Hunter2007}
{Hunter}, J.~D. 2007, Computing in Science and Engineering, 9, 90

\bibitem[{{Karpov} {et~al.}(2025){Karpov}, {Malkov}, \& {Avdeeva}}]{Karpov2025}
{Karpov}, S., {Malkov}, O., \& {Avdeeva}, A. 2025, \aap, 695, A195

\bibitem[{{Kiman} {et~al.}(2026){Kiman}, {Bowler}, {Xuan}, {Mayer}, \&
  {Finley}}]{Kiman2026}
{Kiman}, R., {Bowler}, B.~P., {Xuan}, J.~W., {Mayer}, N., \& {Finley}, C. 2026,
  arXiv e-prints [\eprint[arXiv]{2608.08908}]

\bibitem[{{Kirkpatrick} {et~al.}(2021{\natexlab{a}}){Kirkpatrick}, {Gelino},
  {Faherty}, {Meisner}, {Caselden}, {Schneider}, {Marocco},
  {et~al.}}]{Kirkpatrick2021}
{Kirkpatrick}, J.~D., {Gelino}, C.~R., {Faherty}, J.~K., {et~al.}
  2021{\natexlab{a}}, \apjs, 253, 7

\bibitem[{{Kirkpatrick} {et~al.}(2021{\natexlab{b}}){Kirkpatrick}, {Marocco},
  {Caselden}, {Meisner}, {et~al.}}]{Kirkpatrick2021b}
{Kirkpatrick}, J.~D., {Marocco}, F., {Caselden}, D., {Meisner}, A.~M., {et~al.}
  2021{\natexlab{b}}, \apjl, 915, L6

\bibitem[{{Kirkpatrick} {et~al.}(2024){Kirkpatrick}, {Marocco}, {Gelino},
  {Raghu}, {Faherty}, {et~al.}}]{Kirkpatrick2024}
{Kirkpatrick}, J.~D., {Marocco}, F., {Gelino}, C.~R., {et~al.} 2024, \apjs,
  271, 55

\bibitem[{{Kirkpatrick} {et~al.}(2011)}]{Kirkpatrick2011}
{Kirkpatrick}, J.~D. {et~al.} 2011, ApJS, 197, 19

\bibitem[{{Kiwy} {et~al.}(2022){Kiwy}, {Faherty}, {Meisner},
  {et~al.}}]{Kiwy2022}
{Kiwy}, F., {Faherty}, J.~K., {Meisner}, A., {et~al.} 2022, AJ, 164, 3

\bibitem[{{Kuchner} {et~al.}(2017){Kuchner}, {Faherty}, {Schneider},
  {et~al.}}]{Kuchner2017}
{Kuchner}, M.~J., {Faherty}, J.~K., {Schneider}, A.~C., {et~al.} 2017, \apjl,
  841, L19

\bibitem[{{Lang}(2014)}]{Lang2014}
{Lang}, D. 2014, AJ, 147, 108

\bibitem[{{Lawrence} {et~al.}(2007){Lawrence}, {Warren}, {Almaini}, {Edge},
  {Hambly}, {Jameson}, {Lucas}, {Casali}, {Adamson}, {Dye},
  {et~al.}}]{Lawrence2007}
{Lawrence}, A., {Warren}, S.~J., {Almaini}, O., {et~al.} 2007, MNRAS, 379, 1599

\bibitem[{{Leggett} {et~al.}(2025){Leggett}, {Phillips}, \&
  {Tremblin}}]{Leggett2025}
{Leggett}, S.~K., {Phillips}, M.~W., \& {Tremblin}, P. 2025, ApJ, 991, 193

\bibitem[{{Liu} {et~al.}(2016){Liu}, {Dupuy}, \& {Allers}}]{Liu2016}
{Liu}, M.~C., {Dupuy}, T.~J., \& {Allers}, K.~N. 2016, ApJ, 833, 96

\bibitem[{{Lodieu} {et~al.}(2022){Lodieu}, {Zapatero Osorio}, {Mart{\'\i}n},
  {Rebolo L{\'o}pez}, \& {Gauza}}]{Lodieu2022}
{Lodieu}, N., {Zapatero Osorio}, M.~R., {Mart{\'\i}n}, E.~L., {Rebolo
  L{\'o}pez}, R., \& {Gauza}, B. 2022, A\&A, 663, A84

\bibitem[{{Luhman}(2014)}]{Luhman2014}
{Luhman}, K.~L. 2014, ApJL, 786, L18

\bibitem[{{Luhman} {et~al.}(2024){Luhman}, {Tremblin}, {Alves de Oliveira},
  {Birkmann}, {et~al.}}]{Luhman2024}
{Luhman}, K.~L., {Tremblin}, P., {Alves de Oliveira}, C., {Birkmann}, S.~M.,
  {et~al.} 2024, \aj, 167, 5

\bibitem[{{Mace} {et~al.}(2013){Mace}, {Kirkpatrick}, {Cushing}, {Gelino},
  {et~al.}}]{Mace2013}
{Mace}, G.~N., {Kirkpatrick}, J.~D., {Cushing}, M.~C., {Gelino}, C.~R.,
  {et~al.} 2013, ApJ, 777, 36

\bibitem[{{Mainzer} {et~al.}(2014)}]{Mainzer2014}
{Mainzer}, A. {et~al.} 2014, ApJ, 792, 30

\bibitem[{{Marocco} {et~al.}(2021){Marocco}, {Eisenhardt}, {Fowler},
  {Kirkpatrick}, {et~al.}}]{Marocco2021}
{Marocco}, F., {Eisenhardt}, P.~R.~M., {Fowler}, J.~W., {Kirkpatrick}, J.~D.,
  {et~al.} 2021, \apjs, 253, 8

\bibitem[{{Marocco} {et~al.}(2024){Marocco}, {Kirkpatrick}, {Schneider},
  {Meisner}, {et~al.}}]{Marocco2024}
{Marocco}, F., {Kirkpatrick}, J.~D., {Schneider}, A.~C., {Meisner}, A.~M.,
  {et~al.} 2024, ApJ, 967, 147

\bibitem[{{McMahon} {et~al.}(2013){McMahon}, {Banerji}, {Gonzalez}, {Koposov},
  {Bejar}, {Lodieu}, {Rebolo}, \& {VHS Collaboration}}]{McMahon2013}
{McMahon}, R.~G., {Banerji}, M., {Gonzalez}, E., {et~al.} 2013, The Messenger,
  154, 35

\bibitem[{{Meisner} {et~al.}(2020{\natexlab{a}}){Meisner}, {Caselden},
  {Kirkpatrick}, {Marocco}, {Gelino}, {Cushing}, {et~al.}}]{Meisner2020a}
{Meisner}, A.~M., {Caselden}, D., {Kirkpatrick}, J.~D., {et~al.}
  2020{\natexlab{a}}, ApJ, 889, 74

\bibitem[{{Meisner} {et~al.}(2026){Meisner}, {Caselden}, {Marocco},
  {Kirkpatrick}, {Gagn{\'e}}, {Schneider}, {et~al.}}]{Meisner2026}
{Meisner}, A.~M., {Caselden}, D., {Marocco}, F., {et~al.} 2026, \aj, 172, 154

\bibitem[{{Meisner} {et~al.}(2020{\natexlab{b}}){Meisner}, {Faherty},
  {Kirkpatrick}, {Schneider}, {et~al.}}]{Meisner2020}
{Meisner}, A.~M., {Faherty}, J.~K., {Kirkpatrick}, J.~D., {Schneider}, A.~C.,
  {et~al.} 2020{\natexlab{b}}, ApJ, 899, 123

\bibitem[{{Meisner} {et~al.}(2021)}]{Meisner2021}
{Meisner}, A.~M. {et~al.} 2021, ApJ, 915, 120

\bibitem[{{Metchev} \& {Hillenbrand}(2009)}]{MetchevHillenbrand2009}
{Metchev}, S.~A. \& {Hillenbrand}, L.~A. 2009, ApJS, 181, 62

\bibitem[{{Mukherjee} {et~al.}(2024){Mukherjee}, {Fortney}, {Morley},
  {et~al.}}]{Mukherjee2024}
{Mukherjee}, S., {Fortney}, J.~J., {Morley}, C.~V., {et~al.} 2024, ApJ, 963, 73

\bibitem[{{Nakajima} {et~al.}(1995){Nakajima}, {Oppenheimer}, {Kulkarni},
  {Golimowski}, {Matthews}, \& {Durrance}}]{Nakajima1995}
{Nakajima}, T., {Oppenheimer}, B.~R., {Kulkarni}, S.~R., {et~al.} 1995, \nat,
  378, 463

\bibitem[{{O'Brien} {et~al.}(2023)}]{OBrien2023}
{O'Brien}, M.~W. {et~al.} 2023, MNRAS, 518, 3055

\bibitem[{{Ochsenbein} {et~al.}(2000){Ochsenbein}, {Bauer}, \&
  {Marcout}}]{Ochsenbein2000}
{Ochsenbein}, F., {Bauer}, P., \& {Marcout}, J. 2000, \aaps, 143, 23

\bibitem[{{Pinfield} {et~al.}(2012){Pinfield}, {Burningham}, {Lodieu},
  {et~al.}}]{Pinfield2012}
{Pinfield}, D.~J., {Burningham}, B., {Lodieu}, N., {et~al.} 2012, MNRAS, 422,
  1922

\bibitem[{{Radigan} {et~al.}(2008){Radigan}, {Lafreni{\`e}re}, {Jayawardhana},
  \& {Doyon}}]{Radigan2008}
{Radigan}, J., {Lafreni{\`e}re}, D., {Jayawardhana}, R., \& {Doyon}, R. 2008,
  ApJ, 689, 471

\bibitem[{{Raghu} {et~al.}(2024){Raghu}, {Kirkpatrick}, {Marocco}, {Gelino},
  {et~al.}}]{Raghu2024}
{Raghu}, Y., {Kirkpatrick}, J.~D., {Marocco}, F., {Gelino}, C.~R., {et~al.}
  2024, \apj, 974, 222

\bibitem[{{Rebolo} {et~al.}(1995){Rebolo}, {Zapatero Osorio}, \&
  {Mart{\'\i}n}}]{Rebolo1995}
{Rebolo}, R., {Zapatero Osorio}, M.~R., \& {Mart{\'\i}n}, E.~L. 1995, \nat,
  377, 129

\bibitem[{{Rothermich} {et~al.}(2024){Rothermich}, {Faherty},
  {et~al.}}]{Rothermich2024}
{Rothermich}, A., {Faherty}, J.~K., {et~al.} 2024, AJ, 167, 253

\bibitem[{{Rustamkulov} {et~al.}(2026){Rustamkulov}, {Kirkpatrick}, {Akeson},
  {et~al.}}]{Rustamkulov2026}
{Rustamkulov}, Z., {Kirkpatrick}, J.~D., {Akeson}, R., {et~al.} 2026, \apj,
  1009, 185

\bibitem[{{Sanghi} {et~al.}(2023)}]{Sanghi2023}
{Sanghi}, A. {et~al.} 2023, ApJ, 959, 63

\bibitem[{{Schneider} {et~al.}(2026){Schneider}, {Kuchner}, {Faherty},
  {Meisner}, {Kirkpatrick}, {Burgasser}, {et~al.}}]{Schneider2026}
{Schneider}, A.~C., {Kuchner}, M.~J., {Faherty}, J.~K., {et~al.} 2026, \aj,
  171, 341

\bibitem[{{Schneider} {et~al.}(2020)}]{Schneider2020}
{Schneider}, A.~C. {et~al.} 2020, ApJ, 898, 77

\bibitem[{{Schneider} {et~al.}(2023)}]{SchneiderUHS2023}
{Schneider}, A.~C. {et~al.} 2023, AJ, 166, 103

\bibitem[{{Schneider} {et~al.}(2024)}]{SchneiderHyades2024}
{Schneider}, A.~C. {et~al.} 2024, AJ, 168, 165

\bibitem[{{Scholz}(2010)}]{Scholz2010}
{Scholz}, R.-D. 2010, A\&A, 515, A92

\bibitem[{{Tinney} {et~al.}(2018){Tinney}, {Kirkpatrick}, {Faherty}, {Mace},
  {Cushing}, {et~al.}}]{Tinney2018}
{Tinney}, C.~G., {Kirkpatrick}, J.~D., {Faherty}, J.~K., {et~al.} 2018, ApJS,
  236, 28

\bibitem[{{Tu} {et~al.}(2026){Tu}, {Wang}, {Huang}, {Chen}, \& {Liu}}]{Tu2026}
{Tu}, Z., {Wang}, S., {Huang}, H., {Chen}, X., \& {Liu}, J. 2026, \apj, 1008,
  250

\bibitem[{{Virtanen} {et~al.}(2020)}]{Virtanen2020}
{Virtanen}, P. {et~al.} 2020, Nature Methods, 17, 261

\bibitem[{{Wenger} {et~al.}(2000){Wenger}, {Ochsenbein}, {Egret},
  {et~al.}}]{Wenger2000}
{Wenger}, M., {Ochsenbein}, F., {Egret}, D., {et~al.} 2000, \aaps, 143, 9

\bibitem[{{Wright} {et~al.}(2010)}]{Wright2010}
{Wright}, E.~L. {et~al.} 2010, AJ, 140, 1868

\bibitem[{{Zheng} \& {Gu}(2026)}]{ZhengGu2026}
{Zheng}, S. \& {Gu}, H. 2026, arXiv e-prints, arXiv:2611.34388

\end{thebibliography}

\begin{appendix}

\section{The physical vetting score}\label{app:score}
\begin{figure*}[!tbp]
\centering
\includegraphics[width=0.95\textwidth]{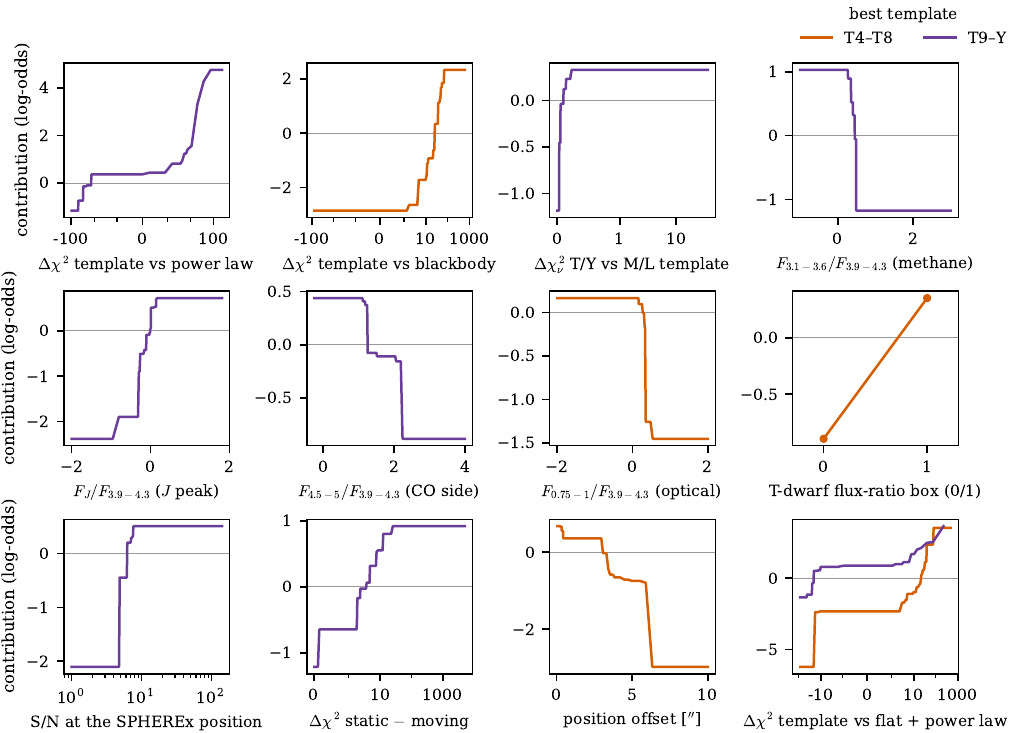}
\caption{The physical vetting score: the contribution of each quantity (log-odds; the score is their sum), for targets with a T4--T8 (orange; five quantities) and a T9--Y (purple; eight quantities) best template. Each contribution is constrained to change monotonically in the physically expected direction. The T-dwarf flux-ratio test is binary: whether the flux ratios at 4.5--5\,\micron, 3.1--3.6\,\micron, $J$ and $K$ fall in the region occupied by T dwarfs.}
\label{fig:vet20}
\end{figure*}

The vetting must reduce about 117\,000 detected targets to a few thousand candidates for inspection without losing real dwarfs systematically. It is a score, the sum of one term per measured quantity (Fig.~\ref{fig:vet20}). Each term is a curve learned from our visual verdicts, but it may only rise or only fall, in the physically expected direction. The score rises when a cold-dwarf template fits the spectrum much better than the shapes of the common contaminants (a blackbody, a power law, or a flat spectrum joined to a power law), when the flux ratios are those of T dwarfs, and when the spectrum shows a methane trough and a $J$ peak; it falls with flux at 0.75--1\,\micron, where a cold dwarf is dark, and with a SPHEREx position far from the predicted one. Because the terms are simply added, each can be read from the figure and checked against physical expectation. We fitted the score separately for targets with T4--T8 and with T9--Y best templates, using the inspected objects as labels and random targets that were never inspected as examples of the population to reject. Neither $W2$ nor Galactic latitude enters the score, so that it does not simply reproduce the selection.

We took three precautions. All numbers are measured out of fold: the objects are split by review page, and each page is scored by a model that never saw it. The false-positive rate is measured, on random targets and on blank sky (Sect.~\ref{sec:val_inj}), not assumed. And the thresholds are set to keep nearly all Detections rather than to minimise contamination, because the inspection removes the contaminants: out of fold, they keep 98\% of the T4--T9 Detections and 94\% of the Y Detections (to the depth covered by the inspection) and send about 0.2\% of random targets to inspection. A classifier that uses all the quantities and their interactions does better only on objects that the inspection removes anyway, so we prefer the score, whose terms can be read.

\paragraph{Checks.} Candidates sent to inspection exclude 36 saturated targets ($W2<10$) and 59 too faint to judge by eye. Near misses just below the thresholds, inspected in full for T4--T5 and in random samples for T9 and Y, contained two T5 Detections at $W2\approx16.4$ and no T9 or Y Detection, so the thresholds are deep enough. The matched filter does not reject outlying exposures, but removing every isolated exposure more than $8\sigma$ from the template changes the S/N of only 2 of the 843 catalogue objects at the time of the test by more than 20\% and moves no Detection below the detection rule. With 80 blank positions, the robust S/N of blank sky has a standard deviation of 1.09, close to the ideal of 1, and $k$ itself is reproducible to 6\% (Sect.~\ref{sec:measurement}). A set of explicit rules, tuned on the bright, well-measured known dwarfs, preceded the score; every candidate they admitted was also inspected, and the inspection decided its class. Two lessons from them shaped the score. Thresholds set on bright dwarfs are too strict for faint ones: the score recovers faint T4--T6 dwarfs with weak methane troughs that the rules rejected (43 Detections and 37 Tentative objects among 295). And motion is a strong argument for a brown dwarf only if the WISE entry is right: several apparent movers were static sources with a wrong catalogue motion or position.

\paragraph{An optical test against extragalactic sources.} Cold brown dwarfs at $W2\approx16$ are extremely faint in the optical, whereas galaxies and AGN of the same $W2$ are often detected and extended. In the Legacy Surveys DR10 \citep{Dey2019} ($\delta\lesssim+32^\circ$), none of the 17 known Y dwarfs with $W2>15$ is detected in $z$, 32\% of the known T9 dwarfs are, always with $z-W2\geq6.6$, and 86\% of the T5--T6.5 dwarfs are, with a median $z-W2$ of 5.5. A Y-type candidate with a $z$ counterpart is therefore not a Y dwarf, and $z-W2<6.5$ also excludes a T9; we use this as a flag, because the detections are close to the survey limit. None of the new Y Detections is detected in $z$ (those north of the DECam footprint only in the shallower Pan-STARRS1). Three T9 Detections have $z-W2<6.5$: CWISE~J021454.06$-$210823.4 and J121143.40+175245.4 ($z-W2=6.4$, faint $z$ detections just below the T9 limit) and J071705.85$-$580823.3 ($z-W2=4.4$, a point source 2\arcsec\ from the position, possibly a companion). We keep them as Detections on their spectra and motions (significant at 3.0, 9.6 and 22$\sigma$, respectively); their types should be checked with higher-resolution spectra.

\section{The excess flux at 3.6--4.1\,\micron}\label{app:excess}
\begin{figure*}[!t]
\centering
\includegraphics[width=\textwidth]{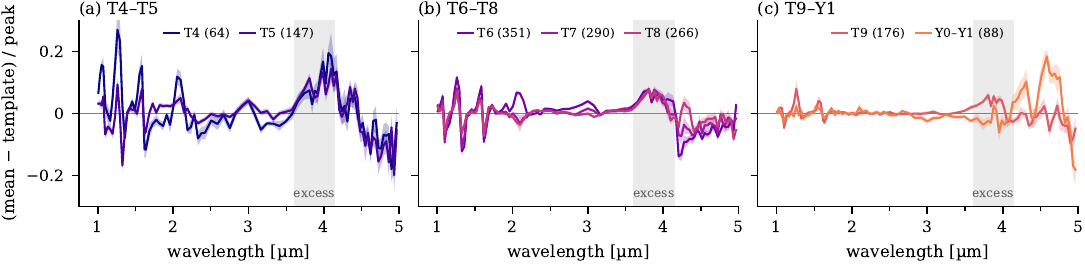}
\caption{The 3.6--4.1\,\micron\ excess (Sect.~\ref{sec:stacks}). Residuals of the mean SPHEREx spectra of Fig.~\ref{fig:stacks} (all Detections of each type; here also T4) from our template of the same type, scaled as in Fig.~\ref{fig:stacks}, in units of the template's 4\,\micron\ peak flux, with the bootstrap 1$\sigma$ band: (a) T4--T5, (b) T6--T8, (c) T9--Y1; the number of dwarfs of each type is given in parentheses. Grey: the 3.6--4.1\,\micron\ excess.}
\label{fig:excess}
\end{figure*}

For the 205 Detections with a T6--T8 type and a peak S/N above 8 (175 known, 23 first confirmations and 7 new), the median excess over the template at 3.9--4.2\,\micron, where it peaks, is 8\% of the peak flux (11\% for T6, 9\% for T7, 4\% for T8), and 41\% of them exceed 10\%; there is a small deficit (3\%) at 4.3--4.5\,\micron. The mean spectra of all Detections show the same pattern at higher significance, with a deficit of up to 10\% at 4.2--4.4\,\micron\ for T6 and T7 (Fig.~\ref{fig:excess}); the T4--T5 means show a larger excess and the T9--Y1 means none.

\section{The online T5--Y brown dwarf atlas}\label{app:atlas}
\begin{figure*}[!tbp]
\centering
\includegraphics[page=1,width=0.86\textwidth]{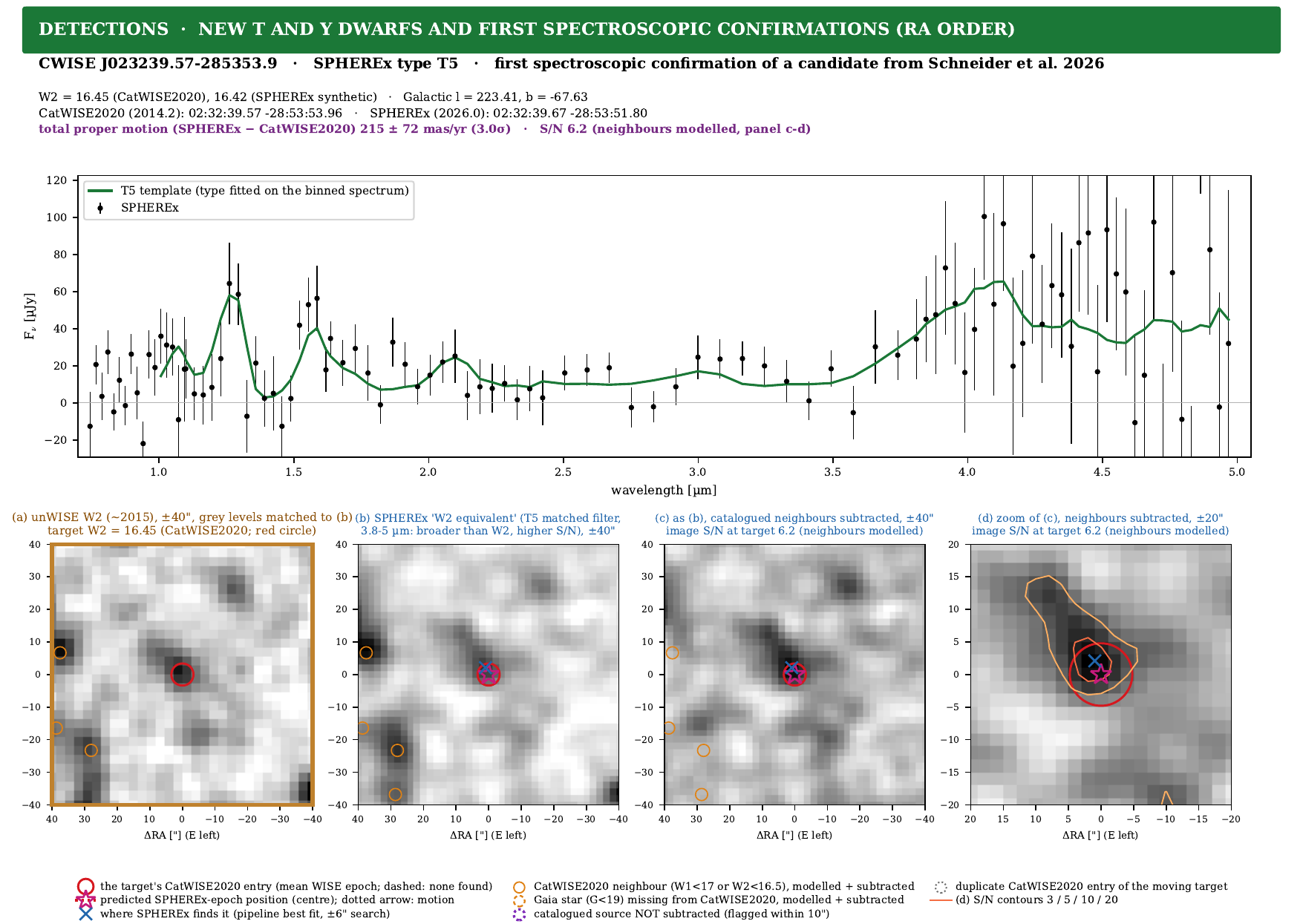}\\[10mm]
\includegraphics[page=2,width=0.86\textwidth]{atlas_pages_v72.pdf}
\caption{Atlas pages of CWISE~J023239.57$-$285353.9 (T5, top), a photometric candidate at $W2=16.45$ confirmed spectroscopically for the first time, and of the new T6 dwarf CWISE~J055347.50+773555.4 (bottom).}
\label{fig:atlas}\label{fig:atlas2}
\end{figure*}

\begin{figure*}[!tbp]
\centering
\includegraphics[page=3,width=0.86\textwidth]{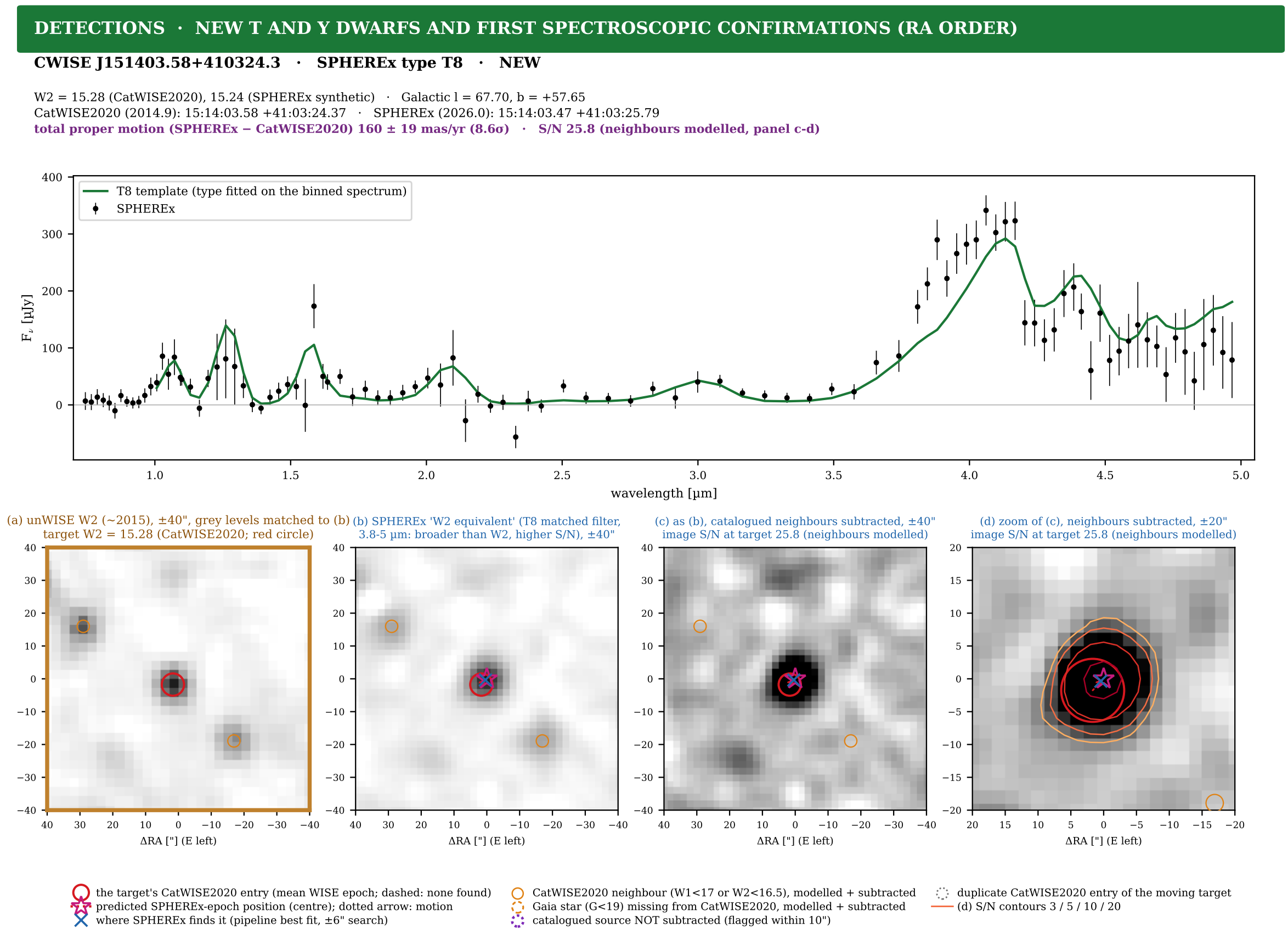}\\[10mm]
\includegraphics[page=4,width=0.86\textwidth]{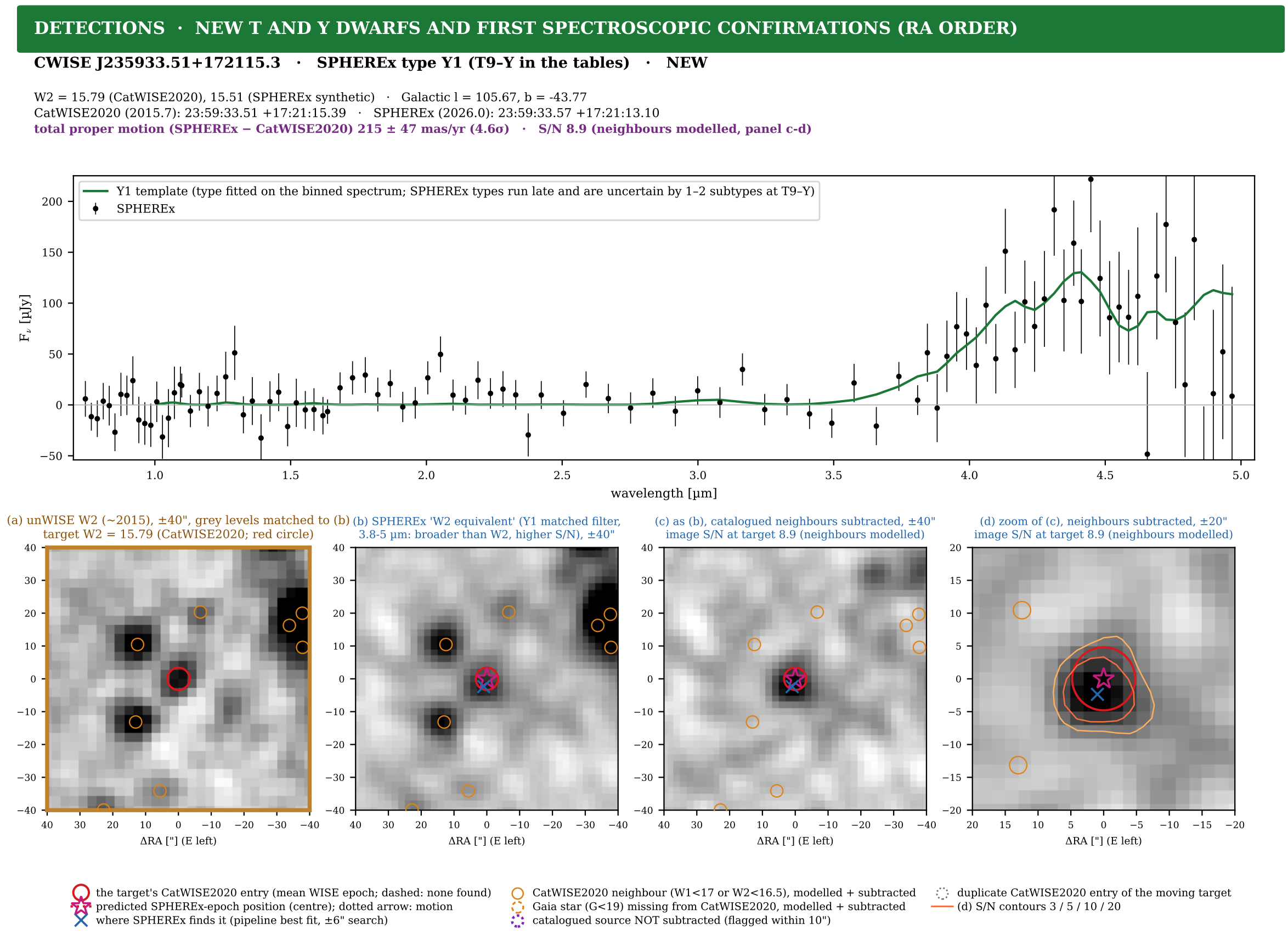}
\caption{Atlas pages of the new T8 dwarf CWISE~J151403.58+410324.3 (top) and the new candidate Y dwarf CWISE~J235933.51+172115.3 (Y1; bottom).}
\label{fig:atlas3}
\end{figure*}

The atlas released with the paper (Sect.~\ref{sec:conclusions}) has an entry for every object, one per page in order of right ascension, in three sections: I, the newly discovered brown dwarfs and the spectroscopically confirmed photometric candidates; II, the previously confirmed brown dwarfs recovered by the search; and III, the tentative objects, kept apart from the others. Figures~\ref{fig:atlas} and \ref{fig:atlas3} reproduce four entries, in their atlas order: a photometric candidate at $W2=16.45$ confirmed here and new T6, T8 and Y1-type dwarfs. Each entry gives the designation and the SPHEREx type fitted on the binned spectrum (for Y types with a reminder that the tables list them as T9--Y, and for T9--Y types a note on their uncertainty; Sect.~\ref{sec:val_known}), and for a confirmed candidate the publication that proposed it; the catalogue and synthetic $W2$ and the Galactic latitude; the CatWISE2020 and SPHEREx positions; and the total motion between them with its significance, and the S/N of the target once its neighbours are subtracted. Below are the binned SPHEREx spectrum with the fitted template and four images: the unWISE $W2$ image; the SPHEREx matched-filter image at 3.8--5\,\micron, the SPHEREx counterpart of the $W2$ image, with a higher S/N because its band is broader; the same image with the modelled catalogued neighbours subtracted; and a zoom of the latter with S/N contours. The first three show the same field, $80\arcsec\times80\arcsec$ on most pages, and all mark the catalogue position (red circle), the position predicted for the SPHEREx epoch (star) and the position measured by SPHEREx (blue cross). Orange circles mark the catalogued neighbours (CatWISE2020, and Gaia DR3 stars missing from it) that the photometry models and subtracts, and dotted purple circles the catalogued sources that it does not subtract; a red line above the spectrum warns that the spectrum may be contaminated, when such a source lies within 10\arcsec\ or the subtracted image shows an unexplained residual source near the target.

\clearpage
\onecolumn
\section{The new T and Y dwarfs, confirmations and tentative objects}\label{app:tables}
Designations are CatWISE2020. $W2$ and $W1-W2$ are from CatWISE2020, $W2_{\rm SX}$ is synthesised from the SPHEREx spectrum, and $J$ (Vega, MKO) is from VHS DR5 \citep{McMahon2013} or UHS DR2 \citep{Dye2018}, or else UKIDSS LAS DR9 \citep{Lawrence2007} or VIKING DR2 \citep{Edge2013}. Objects are grouped by the class of the best template in the detection fit (as in Table~\ref{tab:headline}; the two marked $\star$ are listed as T9 from their $J-W2$); SpT is the type refined on the binned spectrum (Sect.~\ref{sec:measurement}), with the range of templates within $\Delta\chi^2\leq1$ when there is more than one, and can lie outside the group; in addition to this statistical range, the types carry a systematic uncertainty of about $\pm0.7$ subtype (scatter against the literature types of the known dwarfs, ours being on average 0.3 subtype later; Sect.~\ref{sec:val_known}). S/N is the template S/N over 0.75--5\,\micron\ divided by the blank-position factor $k$ (Sect.~\ref{sec:measurement}); objects with S/N$<6$ are detected through their robust S/N at 3.8--5\,\micron\ ($\geq4.5$; Sect.~\ref{sec:vetting}) or kept as Detections by the inspection (Sect.~\ref{sec:vetting}). $\mu_\alpha^*$ and $\mu_\delta$ are the motions between the CatWISE2020 position, at the mean epoch of its WISE observations (2014.8--2015.9), and the SPHEREx position (2025--26), with uncertainty $\sigma_\mu$ per axis, listed only when the motion is significant at $\geq3\sigma$ (otherwise --, with the upper limit on $v_{\rm tan}$). $d$ is the spectrophotometric distance, $v_{\rm tan}$ the tangential velocity (when the motion is not significant, an upper limit from the motion plus 3$\sigma_\mu$) and $T_{\rm eff}$ the Elf Owl temperature. For the confirmed candidates (Table~\ref{tab:fullconf}), Ref. gives the lists that proposed them: B24 \citet{Brooks2024}, C19 \citet{CarneroRosell2019}, D23 \citet{DalPonte2023}, G12 \citet{Griffith2012}, K25 \citet{Karpov2025}, S26 \citet{Schneider2026}; Sim: candidates listed in SIMBAD. Machine-readable versions are available at the CDS.
{\small
\setlength{\tabcolsep}{3.2pt}
\setlength{\LTcapwidth}{\textwidth}
\begin{longtable}{lccccccccccccc}
\caption{New T and Y dwarfs (Detections) from the SPHEREx search. Type: SPHEREx type, i.e. the best-fitting template and the range of templates compatible with the spectrum; at T9--Y it runs one to two subtypes late with respect to the near-infrared scale, so objects of SPHEREx type Y are listed as T9--Y (Sect.~\ref{sec:val_known}). $^{\star}$T9 from its $J-W2$ colour; the best-fitting template is Y0 or Y1 (Sect.~\ref{sec:props}).}\label{tab:fullnew}\\
\toprule
Designation (CWISE) & \multicolumn{1}{c}{$W2$} & \multicolumn{1}{c}{$W2_{\rm SX}$} & \multicolumn{1}{c}{$W1{-}W2$} & \multicolumn{1}{c}{$J$} & \multicolumn{1}{c}{$J{-}W2$} & \multicolumn{1}{c}{SpT} & \multicolumn{1}{c}{S/N} & \multicolumn{1}{c}{$\mu_{\alpha}^{*}$} & \multicolumn{1}{c}{$\mu_{\delta}$} & \multicolumn{1}{c}{$\sigma_\mu$} & \multicolumn{1}{c}{$d$} & \multicolumn{1}{c}{$v_{\rm tan}$} & \multicolumn{1}{c}{$T_{\rm eff}$} \\
 & \multicolumn{5}{c}{(mag)} & & & \multicolumn{3}{c}{(mas\,yr$^{-1}$)} & \multicolumn{1}{c}{(pc)} & \multicolumn{1}{c}{(km\,s$^{-1}$)} & \multicolumn{1}{c}{(K)} \\
\midrule\endfirsthead
\caption{continued.}\\ \toprule
Designation (CWISE) & \multicolumn{1}{c}{$W2$} & \multicolumn{1}{c}{$W2_{\rm SX}$} & \multicolumn{1}{c}{$W1{-}W2$} & \multicolumn{1}{c}{$J$} & \multicolumn{1}{c}{$J{-}W2$} & \multicolumn{1}{c}{SpT} & \multicolumn{1}{c}{S/N} & \multicolumn{1}{c}{$\mu_{\alpha}^{*}$} & \multicolumn{1}{c}{$\mu_{\delta}$} & \multicolumn{1}{c}{$\sigma_\mu$} & \multicolumn{1}{c}{$d$} & \multicolumn{1}{c}{$v_{\rm tan}$} & \multicolumn{1}{c}{$T_{\rm eff}$} \\
 & \multicolumn{5}{c}{(mag)} & & & \multicolumn{3}{c}{(mas\,yr$^{-1}$)} & \multicolumn{1}{c}{(pc)} & \multicolumn{1}{c}{(km\,s$^{-1}$)} & \multicolumn{1}{c}{(K)} \\
\midrule\endhead
\bottomrule\endfoot
\multicolumn{14}{l}{\textit{T4 (best fit) (27)}}\\
J002333.52+215134.6 & 16.27 & 17.15 & 1.92 & \multicolumn{1}{c}{--} & \multicolumn{1}{c}{--} & T2--T4 & 5 & \multicolumn{1}{c}{--} & \multicolumn{1}{c}{--} & 90 & 52 & $<$110 & 1409 \\
J020439.78+462524.2 & 16.32 & 16.72 & 1.85 & \multicolumn{1}{c}{--} & \multicolumn{1}{c}{--} & T4 & 9 & \multicolumn{1}{c}{--} & \multicolumn{1}{c}{--} & 65 & 53 & $<$94 & 1330 \\
J040510.40$-$373351.3 & 16.36 & 16.59 & 1.70 & 18.33 & 1.96 & T4 & 7 & \multicolumn{1}{c}{--} & \multicolumn{1}{c}{--} & 54 & 54 & $<$82 & 1131 \\
J042927.33+181009.8 & 16.11 & 16.07 & 1.53 & \multicolumn{1}{c}{--} & \multicolumn{1}{c}{--} & T4 & 17 & \multicolumn{1}{c}{--} & \multicolumn{1}{c}{--} & 58 & 48 & $<$67 & 1281 \\
J044819.02+091444.3 & 16.21 & 15.83 & 1.66 & \multicolumn{1}{c}{--} & \multicolumn{1}{c}{--} & T4 & 9 & \multicolumn{1}{c}{--} & \multicolumn{1}{c}{--} & 92 & 51 & $<$102 & 1273 \\
J053411.81$-$152349.1 & 16.36 & 16.72 & 1.66 & 18.19 & 1.83 & T4--T6 & 9 & 54 & $-$289 & 72 & 54 & 76 & 1245 \\
J063137.18$-$351544.5 & 16.03 & 16.75 & 1.99 & \multicolumn{1}{c}{--} & \multicolumn{1}{c}{--} & T4 & 9 & $-$162 & 197 & 43 & 46 & 56 & 1183 \\
\end{longtable}

\setlength{\tabcolsep}{3.2pt}
\setlength{\LTcapwidth}{\textwidth}
\begin{longtable}{lcccccccccccccc}
\caption{Published photometric or motion candidates confirmed spectroscopically for the first time.}\label{tab:fullconf}\\
\toprule
Designation (CWISE) & \multicolumn{1}{c}{$W2$} & \multicolumn{1}{c}{$W2_{\rm SX}$} & \multicolumn{1}{c}{$W1{-}W2$} & \multicolumn{1}{c}{$J$} & \multicolumn{1}{c}{$J{-}W2$} & \multicolumn{1}{c}{SpT} & \multicolumn{1}{c}{S/N} & \multicolumn{1}{c}{$\mu_{\alpha}^{*}$} & \multicolumn{1}{c}{$\mu_{\delta}$} & \multicolumn{1}{c}{$\sigma_\mu$} & \multicolumn{1}{c}{$d$} & \multicolumn{1}{c}{$v_{\rm tan}$} & \multicolumn{1}{c}{$T_{\rm eff}$} & \multicolumn{1}{c}{Ref.} \\
 & \multicolumn{5}{c}{(mag)} & & & \multicolumn{3}{c}{(mas\,yr$^{-1}$)} & \multicolumn{1}{c}{(pc)} & \multicolumn{1}{c}{(km\,s$^{-1}$)} & \multicolumn{1}{c}{(K)} &  \\
\midrule\endfirsthead
\caption{continued.}\\ \toprule
Designation (CWISE) & \multicolumn{1}{c}{$W2$} & \multicolumn{1}{c}{$W2_{\rm SX}$} & \multicolumn{1}{c}{$W1{-}W2$} & \multicolumn{1}{c}{$J$} & \multicolumn{1}{c}{$J{-}W2$} & \multicolumn{1}{c}{SpT} & \multicolumn{1}{c}{S/N} & \multicolumn{1}{c}{$\mu_{\alpha}^{*}$} & \multicolumn{1}{c}{$\mu_{\delta}$} & \multicolumn{1}{c}{$\sigma_\mu$} & \multicolumn{1}{c}{$d$} & \multicolumn{1}{c}{$v_{\rm tan}$} & \multicolumn{1}{c}{$T_{\rm eff}$} & \multicolumn{1}{c}{Ref.} \\
 & \multicolumn{5}{c}{(mag)} & & & \multicolumn{3}{c}{(mas\,yr$^{-1}$)} & \multicolumn{1}{c}{(pc)} & \multicolumn{1}{c}{(km\,s$^{-1}$)} & \multicolumn{1}{c}{(K)} &  \\
\midrule\endhead
\bottomrule\endfoot
\multicolumn{15}{l}{\textit{T4 (best fit) (13)}}\\
J000149.45$-$061336.7 & 15.17 & 15.02 & 1.56 & 16.84 & 1.68 & T4 & 17 & $-$58 & $-$71 & 20 & 31 & 14 & 1137 & D23 \\
J023746.91$-$543658.3 & 15.86 & 16.25 & 1.54 & 17.38 & 1.53 & T4 & 13 & 129 & 26 & 32 & 43 & 27 & 1155 & C19 \\
J033055.97$-$642256.1 & 16.02 & 16.10 & 1.52 & 17.66 & 1.63 & T4--T6 & 15 & 205 & 143 & 56 & 46 & 55 & 1191 & S26 \\
J045126.38$-$544156.8 & 16.30 & 16.45 & 1.61 & 17.85 & 1.55 & T4 & 11 & 311 & 192 & 28 & 53 & 91 & 1205 & D23 \\
J071250.77$-$804800.9 & 15.25 & 14.51 & 1.68 & 16.84 & 1.59 & T4 & 30 & 103 & 484 & 23 & 32 & 76 & 1499 & K25 \\
J095857.97$-$300718.2 & 15.74 & 15.70 & 1.63 & 18.24 & 2.50 & T4 & 10 & $-$201 & 149 & 47 & 41 & 48 & 1245 & Sim \\
J101410.48$-$723625.2 & 15.58 & 15.44 & 1.74 & 17.40 & 1.81 & T4 & 16 & \multicolumn{1}{c}{--} & \multicolumn{1}{c}{--} & 26 & 38 & $<$24 & 1205 & G12 \\
\end{longtable}

\setlength{\tabcolsep}{3.2pt}
\setlength{\LTcapwidth}{\textwidth}
\begin{longtable}{lcccccccccc}
\caption{Tentative T and Y dwarfs, kept separate from the Detections and not used in any statistic; their types are uncertain, so no distance, tangential velocity or temperature is given.}\label{tab:fulltent}\\
\toprule
Designation (CWISE) & \multicolumn{1}{c}{$W2$} & \multicolumn{1}{c}{$W2_{\rm SX}$} & \multicolumn{1}{c}{$W1{-}W2$} & \multicolumn{1}{c}{$J$} & \multicolumn{1}{c}{$J{-}W2$} & \multicolumn{1}{c}{SpT} & \multicolumn{1}{c}{S/N} & \multicolumn{1}{c}{$\mu_{\alpha}^{*}$} & \multicolumn{1}{c}{$\mu_{\delta}$} & \multicolumn{1}{c}{$\sigma_\mu$} \\
 & \multicolumn{5}{c}{(mag)} & & & \multicolumn{3}{c}{(mas\,yr$^{-1}$)} \\
\midrule\endfirsthead
\caption{continued.}\\ \toprule
Designation (CWISE) & \multicolumn{1}{c}{$W2$} & \multicolumn{1}{c}{$W2_{\rm SX}$} & \multicolumn{1}{c}{$W1{-}W2$} & \multicolumn{1}{c}{$J$} & \multicolumn{1}{c}{$J{-}W2$} & \multicolumn{1}{c}{SpT} & \multicolumn{1}{c}{S/N} & \multicolumn{1}{c}{$\mu_{\alpha}^{*}$} & \multicolumn{1}{c}{$\mu_{\delta}$} & \multicolumn{1}{c}{$\sigma_\mu$} \\
 & \multicolumn{5}{c}{(mag)} & & & \multicolumn{3}{c}{(mas\,yr$^{-1}$)} \\
\midrule\endhead
\bottomrule\endfoot
\multicolumn{11}{l}{\textit{T4 (best fit) (12)}}\\
J044026.12+182328.5 & 16.29 & 16.99 & 2.06 & \multicolumn{1}{c}{--} & \multicolumn{1}{c}{--} & T4 & 6 & \multicolumn{1}{c}{--} & \multicolumn{1}{c}{--} & 124 \\
J063652.72+451024.9 & 16.11 & 16.07 & 2.41 & \multicolumn{1}{c}{--} & \multicolumn{1}{c}{--} & T4 & 6 & \multicolumn{1}{c}{--} & \multicolumn{1}{c}{--} & 101 \\
J064628.29$-$401834.9 & 16.31 & 16.49 & 2.55 & \multicolumn{1}{c}{--} & \multicolumn{1}{c}{--} & T4 & 4 & \multicolumn{1}{c}{--} & \multicolumn{1}{c}{--} & 81 \\
J070539.37+235120.6 & 16.21 & 16.94 & 2.29 & \multicolumn{1}{c}{--} & \multicolumn{1}{c}{--} & T4 & 5 & $-$44 & $-$324 & 85 \\
J101301.79+303824.3 & 16.48 & 17.26 & 1.62 & \multicolumn{1}{c}{--} & \multicolumn{1}{c}{--} & T4 & 7 & \multicolumn{1}{c}{--} & \multicolumn{1}{c}{--} & 114 \\
J102750.89$-$410151.9 & 16.14 & 15.51 & 2.95 & 18.06 & 1.91 & T4 & 15 & $-$3 & $-$85 & 24 \\
J113100.97$-$481650.2 & 16.23 & 16.98 & 1.66 & 18.58 & 2.35 & T4--T5 & 5 & 48 & 132 & 40 \\
\end{longtable}

}

\end{appendix}

\label{lastpage}
\end{document}